\documentclass[secnumarabic,amssymb, nobibnotes, aps, prd]{revtex4-2}
\usepackage{algorithmic}
\usepackage{algorithm}
\usepackage{amsmath}
\usepackage{txfonts}
\usepackage{color}
\usepackage{graphicx}  
\usepackage{here}
\usepackage{braket}
\usepackage{bm}
\usepackage[OT1]{fontenc} 

\begin{document}
\title{Bayesian Inference of Absorption Spectra Based on Binomial Distribution}

\author{Tomohiro Nabika$^1$, Kenji Nagata$^2$, Shun Katakami$^1$, Masaichiro Mizumaki$^3$,\\and Masato Okada$^1$}
\affiliation{$^1$Graduate School of Frontier Sciences, The University of Tokyo, Kashiwa, Chiba 277-8561, Japan, \\
$^2$Research and Services Division of Materials Data and Integrated System, National Institute for Materials Science,Tsukuba, Ibaraki 305-0047, Japan,\\
$^3$Faculty of Science, Course for Physical Sciences, Kumamoto University, Kumamoto, Kumamoto 860-8555, Japan.} 


\begin{abstract}
In this paper, we propose a Bayesian spectral deconvolution method for absorption spectra.
In conventional analysis, the noise mechanism of absorption spectral data is never considered appropriately. In that analysis, the least-squares method, which assumes Gaussian noise from the perspective of Bayesian statistics, is frequently used.
 Since Bayesian inference is possible by introducing an appropriate noise model for the data, we consider the absorption process of a single photon to be a Bernoulli trial and develop a Bayesian spectral deconvolution method based on binomial distribution.
  We have evaluated our method on artificial data under several conditions by numerical experiments. 
  The results show that our method not only allows us to estimate parameters with high accuracy from absorption spectral data, but also to infer them even from absorption spectral data with large absorption rates where the spectral structure is flattened, which was previously impossible to analyze.
\end{abstract}
\maketitle
    
\section{Introduction}
Absorption spectra are frequently measured and analyzed in many fields of natural science to investigate the properties of materials.
For example, in condensed matter science, the atomic state is studied by X-ray absorption spectroscopy (XAS), a method to determine the energy levels of electrons from the absorption rate of X-rays in materials \cite{XAS}. In chemistry and planetary science, near-infrared spectroscopy (NIR spectroscopy), which obtains information about the chemical structure from near-infrared absorption, enables the identification of the chemical substances that the measurement target contains \cite{NIR}.
In particular, NIR spectroscopy has been applied to remote sensing, a measurement in which the observation target cannot be touched, to study the atmospheric conditions \cite{Pfeilsticker2003} and the geological structure of planets \cite{Pieters1982}.

In the analysis of absorption spectra, the absorption rate, which is the ratio of the incident amount of photons to the absorbed amount of photons, is often handled.
Since the absorption rate is the ratio of two quantities measured stochastically, it is difficult to discuss as a simple stochastic model, and noise mechanisms in absorption spectrum data are never considered appropriately. Hence, the absorption spectral data are analyzed by least square error fitting to obtain a curve of the absorption spectrum showing the relationship between the optical wavelength and the amount of absorption \cite{Sunshine1998}.
Least square error fitting means that the noise mechanism of the data follows Gaussian noise, according to the interpretation in Bayesian inference.
However, there is no evidence that the noise in the measured data of photon incidence or photon absorption follows Gaussian noise, much less that the noise in the ratio of these two data follows normal noise.
It is important to consider the noise mechanism carefully because Bayesian inference requires an appropriate noise model for the data. Forcing an incorrect noise mechanism into a Bayesian model will lead to incorrect inference. It is known that introducing appropriate noise mechanisms in the data into the analysis improves the performance of the analysis. Katakami et al. showed that an appropriate noise model can improve estimation performance and enable estimation of physical quantities even from measurement data with low signal-to-noise ratios \cite{Katakami2022}. Therefore, in this paper, we consider the absorption process of a single photon to be a Bernoulli trial and assume that the noise in the absorption spectrum data follows a binomial distribution noise.
\par
Bayesian spectral deconvolution method is effective in the analysis of spectra and has been applied to various studies \cite{Spectral_2,Nagata2012,Nagata2019,Machida2021,Kashiwamura2022}.
For example, in X-ray photoelectron spectroscopy(XPS), deconvolution of spectral data into basic peak functions is performed by Bayesian inference and the parameter of every peaks, the number of peaks and the compound ratio of the measurement target are estimated with statistical reliability assessment \cite{Machida2021}.
However, the noise models of measurement data that can be handled by Bayesian spectral deconvolution method proposed so far are limited to Gaussian noise \cite{Nagata2012} or Poisson noise \cite{Nagata2019}. In other words, it is not possible to directly apply existing Bayesian spectral deconvolution method to the analysis of absorption spectral data.
In this study, we propose a Bayesian spectral deconvolution method for absorption spectral data based on the binomial distribution.

\begin{figure}[h]
    \centering
    \includegraphics[width = 8.0cm]{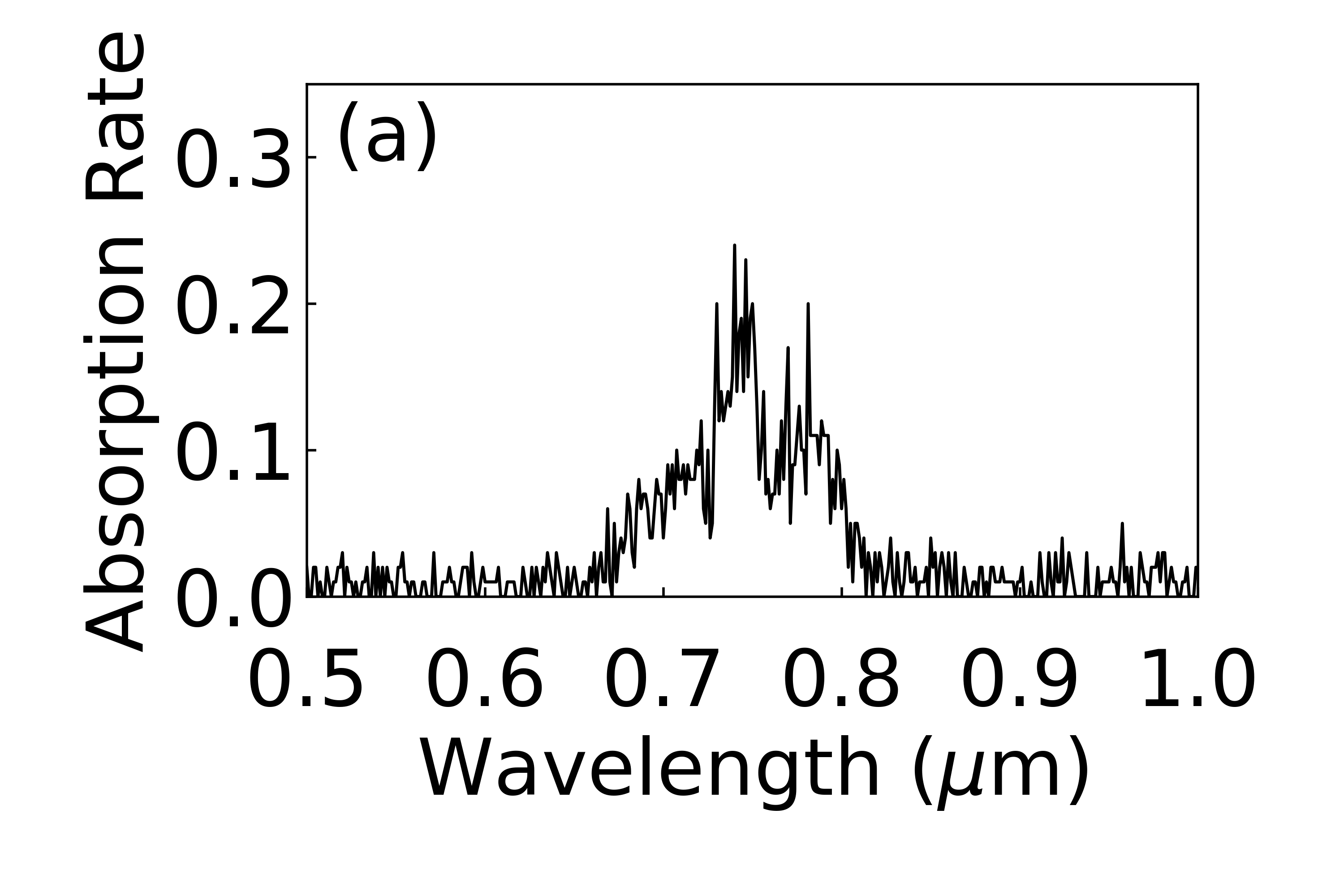}
    \hspace{-1.0cm}
    \includegraphics[width = 8.0cm]{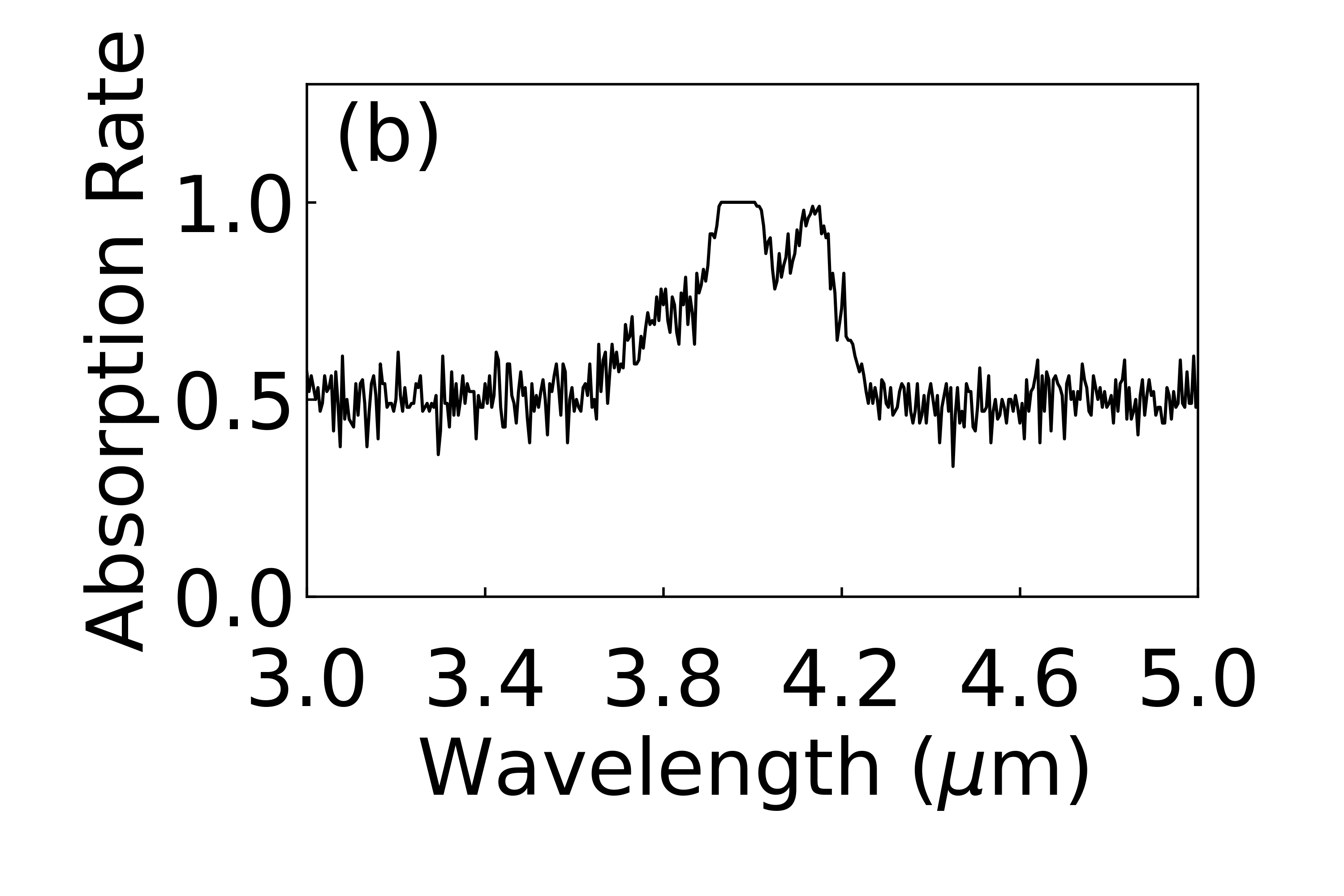}
    \caption{Sketches of absorption spectra when the incident photons per measurement point $N = 100$. (a) Absorption rate is small enough to observe the peak structures. (b) Absorption rate is so large that the spectral structure is flattened and peak structures cannot be observed. Data similar to (b) are obtained in planetary science where the target substance cannot be adjusted and spectra for a wide range of wavelengths for the same substance is observed \cite{Goody1995}. The large background function in (b) indicates the small density of the focused material.}
    \label{earth_atmosphere}
\end{figure}

Our method is not only an extension of the spectral deconvolution to absorption spectra but also expected to expand the analysis range of absorption spectroscopy to include objects that could not be analyzed. 
Figure \ref{earth_atmosphere} shows the sketch of absorption spectra. In the conventional analysis, spectral analysis was performed only in range where absorption rate is small enough to observe the peak structures, and the spectral data in the range where the absorption rate is too large to observe peak structures have been excluded from the analysis. 
In the field of materials science, the absorption rate can be adjusted to be small. However, in the field of planetary science, the target substance cannot be adjusted and the absorption rate can be near or above unity absorption especially when the range of wavelength is large.
For example, the peak structure of absorption rate can be flattened in the analysis of the earth's atmosphere \cite{Goody1995}.
We show that binomial distribution noise model enables Bayesian inference to work properly even in the range where the peak structures cannot be observed through artificial data analysis. 
 \par
In this paper, we evaluate the effectiveness of the method by analyzing artificial data of absorption spectra.
We performed a Bayesian spectral deconvolution of data generated based on binomial distribution for the cases where the absorption rate is sufficiently small that the peak structures are observable and too large that the peak structures are not observable. 
We found that Bayesian spectral deconvolution can be performed even for the data with low signal-to-noise ratios, and the data in which the peak structure cannot be observed.\par

\color{black}

The structure of this paper is as follows. In Sect. 2, we formulate a probabilistic model of spectral measurement of ratio such as absorption spectra, etc. In Sect. 3, we describe Bayesian spectral deconvolution based on binomial distribution. In Sect. 4, we evaluate the performance of the methods described in Sect. 3 with artificial data. In Sect. 5, we conclude this paper and discuss future work.

\section{Spectral Measurement Based on Binomial Distribution}
In this section, we describe a probabilistic model of the measurement of absorption spectra. 
In Sect. 2.1, we formulate a probabilistic model of the process of generating data that are generally measured as ratios, and in Sect. 2.2, we explain the specific formulation for absorption spectra.
\subsection{Data Generation Based on Binomial Distribution}
Suppose that event A occurs with a certain probability $\alpha$ and follows Bernoulli distribution. When this event is repeated $N$ times independently, the number $n$ of times event A occurs follows a binomial distribution as follows:
\begin{align}
    p(n) = \binom{N}{n}\ \alpha^n(1-\alpha)^{N - n},
\end{align}
 where $\binom{N}{n} = \frac{N!}{n!(N-n)!}$ \cite{PRML}.
If the proportion is given as a function, such that $\alpha = \min(f(x),1)$, then the number $n$ of events that have occurred corresponding to $x,N$ follows a binomial distribution:
\begin{align}
    p(n|x,N) = \binom{N}{n}\ \min(f(x),1)^{n}(1-\min(f(x),1))^{N - n}.
\end{align}
This formulation can be applied to measurements of absorption rate of materials, by setting $N$ to be the number of incident photons and $n$ to be the number of absorbed photons, because each photon is determined to be absorbed at a certain absorption rate.
\subsection{Absorption Spectral Measurement Based on Binomial Distribution}
In this subsection, we describe the formulation of the probabilistic model when an absorption spectrum can be represented by a linear sum of basis functions.
Let the probability $\alpha$ be the absorption rate, $N$ be the number of photons incident on a material, \ $x$ be the energy of the spectrum, $K$ be the number of peaks, $w = \{a_k, \mu_k, \sigma_k\}_{k = 1}^K$ be the parameters of basic functions\ 
($a_k,\mu_k,\sigma_k$ indicates the peak intensity, the peak position, the peak width respectively), 
$v$ be the parameter of background functions, $\phi$ be the basis function, and $\theta = \{w,v\}$ be the parameter for the absorption rate. We then define $G(x; w, K)$ as
\begin{align}
    G(x; w, K) := \sum_{k = 1}^{K} a_k\phi(x;\mu_k,\sigma_k).
\end{align} 
If the background function is $B(x;\theta,K)$, the absorption rate $f(x;\theta,K)$ can be formulated as
\begin{align}
    f(x;\theta,K) := G(x;w,K) + B(x;\theta,K).
\end{align}
Here, the basis function $\phi$ can be Gaussian, Lorentzian, Voigt function, etc., and the background function can be the constant model $B(x;\theta,K) = B$ ($v = \{B\}$), etc.\par
Let $N$ be the number of incident photons. We can assume that the number of absorbed photons follows a binomial distribution as follows:
\begin{align}
    p(n|x,N,\theta,K) = \binom{N}{n}\ \min(f(x;\theta,K),1)^{n}(1-\min(f(x;\theta,K),1))^{N - n}.
\end{align}
\section{Bayesian Spectral Deconvolution Based on Binomial Distribution}
In this section, we describe Bayesian spectral deconvolution when the data are generated according to a binomial distribution. 
In Sect. 3.1, we formulate Bayesian spectral deconvolution based on a binomial distribution. 
In Sect. 3.2, 
we explain how they are computed using Markov chain Monte Carlo (MCMC) methods.

\subsection{Bayesian Estimation Based on Binomial Distribution}
Let $M$ be the number of data points and $D = \{x_i,N_i,n_i\}_{i=1}^M$ be the data set (where $x_i$ is the energy of incident photons, $N_i$ is the number of incident photons, and $n_i$ is the number of absorbed photons).
Then, the probability distribution of the data set $D$ is as follows:
\begin{align}
    p(D|\theta,K) &= \prod_{i=1}^M p(n_i|x_i,N_i,K) := \exp(-ME(\theta,K)), \\
    E(\theta,K) &= \frac{1}{M}\sum_{i=1}^M \Biggl\{ n_i \log(\min(f(x_i;\theta,K),1))
    + (N_i - n_i)\log(1-\min(f(x_i;\theta,K),1)) + \log\binom{N_i}{n_i}\Biggr\}.
\end{align}
Assuming that the number of peaks $K$ follows the prior distribution $p(K)$ and the parameter $\theta$ follows the prior distribution $p(\theta|K)$, the joint distribution is
\begin{align}
    p(D,\theta,K) = p(D|\theta,K)p(\theta|K)p(K).
\end{align}
Here, on the basis of Bayes' theorem \cite{BayesianInference}, the posterior probability of the parameter $\theta$ is as follows:
\begin{align}
    p(\theta|D,K) &= \frac{p(D,\theta,K)}{\int p(D,\theta,K)\textup{d}\theta} \\
                  &= \frac{1}{Z(K)}\exp\left(-ME(\theta,K)\right)p(\theta,K), \\
    Z(K) &= \int \exp\left( -ME(\theta,K)\right) p(\theta|K)\textup{d}\theta.        
\end{align}
Moreover, the posterior probability of $K$ is as follows:
\begin{align}
    p(K|D) &= \frac{\int p(D,\theta,K)\textup{d}\theta}{\sum_K \int p(D,\theta,K)\textup{d}\theta} \\
    \label{P(K|D)}
                  &= \frac{p(K)}{\bar{Z}}\exp(-F(K)), \\
    F(K) &= -\log \int \exp\left( -ME(\theta,K)\right) p(\theta|K)\textup{d}\theta, \\
    \bar{Z} &= \sum_K \int \exp \left( -M E(\theta,K)\right) p(\theta|K)p(K)\textup{d}\theta.      
\end{align}
In this study, $p(K)$ is assumed to be uniformly distributed.
In the numerical experiments, $K$ is estimated by maxmizing the posterior probability $p(K|D)$.
The parameter $\theta$ is estimated by maxmizing the posterior probability $p(\theta|D,K)$ with a given $K$.
\subsection{Bayesian Spectral Deconvolution using MCMC}
It is impossible to calculate $p(\theta|D,K)$ and $p(K|D)$ analytically since the calculations of $Z(K)$ requires a high-dimensional integration.
Therefore, we obtain samples following $p(\theta|D,K)$ and calculate $p(K|D)$ numerically by the exchange Monte Carlo (EMC) method. The EMC method is one of the
Markov chain Monte Carlo (MCMC) methods and useful for obtaining samples that follow a high-dimensional probability distribution and for performing high-dimensional integrations \cite{Nagata2012,Hukushima1996}.
We prepared the replicated probability distribution $\{p_{\beta_l}(\theta|D,K)\}_{l = 1}^L$ with the inverse temperature parameter $0 < \beta_1 < \cdots < \beta_L = 1$ as follows:
\begin{align}
    p_{\beta}(\theta|D,K) \propto \exp(-M\beta E(\theta,K))p(\theta|K).
\end{align}
In the EMC method, we can obtain samples following $\{p_{\beta_l}(\theta|D,K) \}_{l = 1}^L$. The specific algorithm is described in Algorithm \ref{EMC}. By focusing on the inverse temperature $\beta_L = 1$, we can obtain samples following the distribution of \ $p(\theta|D,K)$ and find $\theta$ where $p(\theta|D,K)$ is maximum.
For the calculation of free energy, we define $z(\beta)$ as
\begin{align}
    z(\beta) = \int \exp(-M\beta E(\theta,K))p(\theta|K)\textup{d}\theta.
\end{align}
Then, $z(0) = 1$ and the free energy $F(K) =  -\log(z(1))$.
$z(1)$ can be computed for the sample following $\{p_{\beta_l}(\theta|D,K)\}_{l = 1}^L$ obtained by EMC by the following deformation:
\begin{align}
    z(1) &= \frac{z(\beta_L)}{z(\beta_{L-1})} \times \cdots \times \frac{z(\beta_2)}{z(\beta_1)} = \prod_{l=1}^{L-1} \frac{z(\beta_{l+1})}{z(\beta_l)} \\
         &= \prod_{l=1}^{L-1} \frac{\int \exp(-M\beta_{l+1})E(\theta,K)p(\theta|K)}{\int \exp(-M\beta_l E(\theta,K))p(\theta|K)} \\
         &= \prod_{l=1}^{L-1} \Braket{\exp(-M(\beta_{l+1} - \beta_{l})E(\theta,K))}_{p_{\beta_l}(\theta|D,K)}.
\end{align}
From Equation(\ref{P(K|D)}), we can calculate the posterior probability $P(K|D)$ from $F(K)$.
\begin{figure}
    \begin{algorithm}[H]
        \caption{Exchange Monte Carlo Method for Bayesian Estimation}
        \label{EMC}
        \begin{algorithmic}[1]
        \REQUIRE Data $D = \{x_i,N_i,n_i\}_{i=1}^M$,\ Number of peaks $K$,\ Probability distribution $P(\theta|K)$,\ Inverse temperatures $\{\beta_l\}_{l = 1}^L$,\ Burn in $T_1$, Sample size $T_2$
        \ENSURE Distribution $\Theta$ = \{$\{\theta_{l,t}\}_{t = T_1+1}^{T_2}$ which follows $p_{\beta_l}(\theta|D,K)\}_{l = 1}^L$
        \STATE Distribution $\Theta$ = \{\}
        \FOR{$l \in \{1,...,L\}$}
            \STATE Generate $\theta_{l}$ with the probability $P(\theta|K)$
        \ENDFOR
        \FOR{$t \in \{1,...,T_2\}$}
            \FOR{$l \in \{1,...,L\}$}
                \STATE Update parameter $\theta_{l}$ using the Metropolis algorithm.
            \ENDFOR
            \FOR{$l \in \{1,...,L-1\}$}
                \STATE Calculate $v = \dfrac{p_{\beta_{l}}(\theta_{l+1}|D,K)p_{\beta_{l+1}}(\theta_{l}|D,K)}{p_{\beta_{l}}(\theta_{l}|D,K)p_{\beta_{l+1}}(\theta_{l+1}|D,K)}$
                \STATE Exchange $\theta_l$ and $\theta_{l+1}$ with the possibility $\min(1,v)$
            \ENDFOR
            \IF{$t > T_1$}
            \STATE Add $\{\theta_l\}_{l=1}^L$ in $\Theta$
            \ENDIF 
        \ENDFOR
        \end{algorithmic}
    \end{algorithm}
\end{figure}

\section{Validation of Our Proposed Method using Artificial Data}
In this section, we describe the validation of the Bayesian spectral deconvolution method based on a binomial distribution using artificial data.
In Sect. 4.1, we describe the problem settings when peak structures can be observed and cannot be observed.
In Sect. 4.2, 
we report the result of the Bayesian spectral deconvolution based on the binomial distribution.

Here, let the number of incident photons for a measurement point $N_i$ be constant $N_i = N$ independent of $x$, the basis function $\phi(x;\mu,\sigma)$ be
\begin{align}
    \phi(x;\mu,\sigma) = \exp\left(-\frac{(x-\mu)^2}{2\sigma^2}\right),
\end{align}
and the background function be $B(x;\theta,K) = B$. 
\subsection{Problem Settings}
In Sect. 4,  we consider two situations where the peak structures are small enough to be observed and the peak structures are too large to be observed.
\subsubsection{Case where peak structures are observable}
In the case where the peak structures are observable, let the number of peaks $K=3$ and the true parameter $\theta^* = \{ \{ a_k^*, \mu_k^*, \sigma_k^* \}_{k=1}^3, B^* \}$ be as follows:
\begin{align}
    \begin{pmatrix}
        a_1^*\\
        a_2^*\\
        a_3^*\\
    \end{pmatrix} 
    = 
    \begin{pmatrix}
        0. 0587\\
        0.1522\\
        0.1183\\
    \end{pmatrix}  ,\ 
    \begin{pmatrix}
        \mu_1^*\\
        \mu_2^*\\
        \mu_3^*\\
    \end{pmatrix}
    = 
    \begin{pmatrix}
        0.7016\\
        0.7426\\
        0.7838\\
    \end{pmatrix},\
    \begin{pmatrix}
        \sigma_1^*\\
        \sigma_2^*\\
        \sigma_3^*\\
    \end{pmatrix}
    = 
    \begin{pmatrix}
        0.01705\\
        0.01375\\
        0.01300\\
    \end{pmatrix},\ 
    B = 0.01.
\end{align}
We adopted the shape of true value from the artificial data from Nagata et al. who generate true parameters from prior distribution and select one parameter which is somewhat difficult to select the number of peaks \cite{Nagata2012}. 
This shape of true value makes it easy to understand the relationship between the signal-to-noise ratio and the probability of success of model selection and parameter estimation. Moreover, this shape can be appeared in real absorption spectra \cite{Chai2016}.\par
\begin{figure}
    \centering
    \includegraphics*[width = .27\columnwidth]{"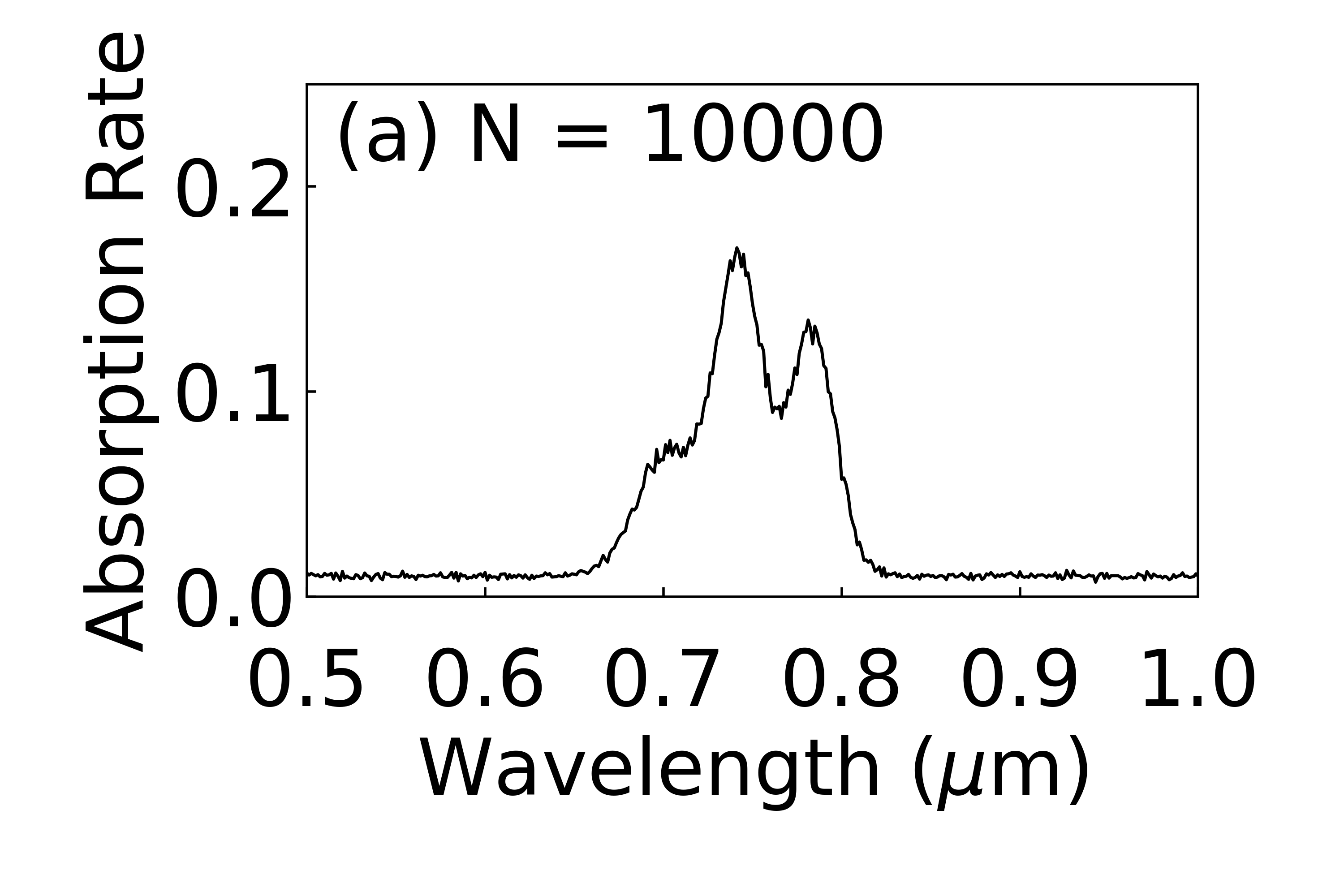"}
    \hspace{-0.7cm}
    \includegraphics*[width = .27\columnwidth]{"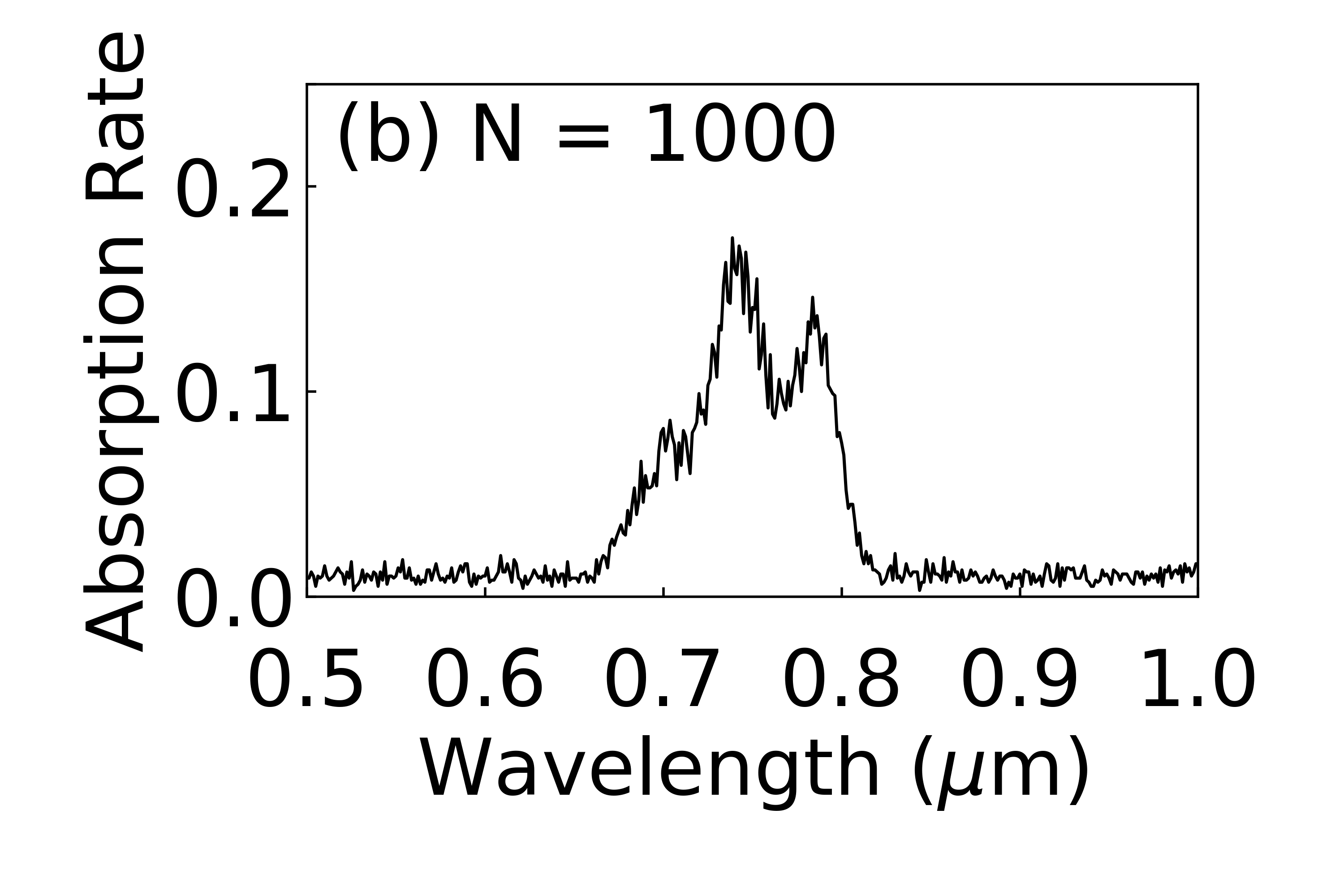"}
    \hspace{-0.7cm}
    \includegraphics*[width = .27\columnwidth]{"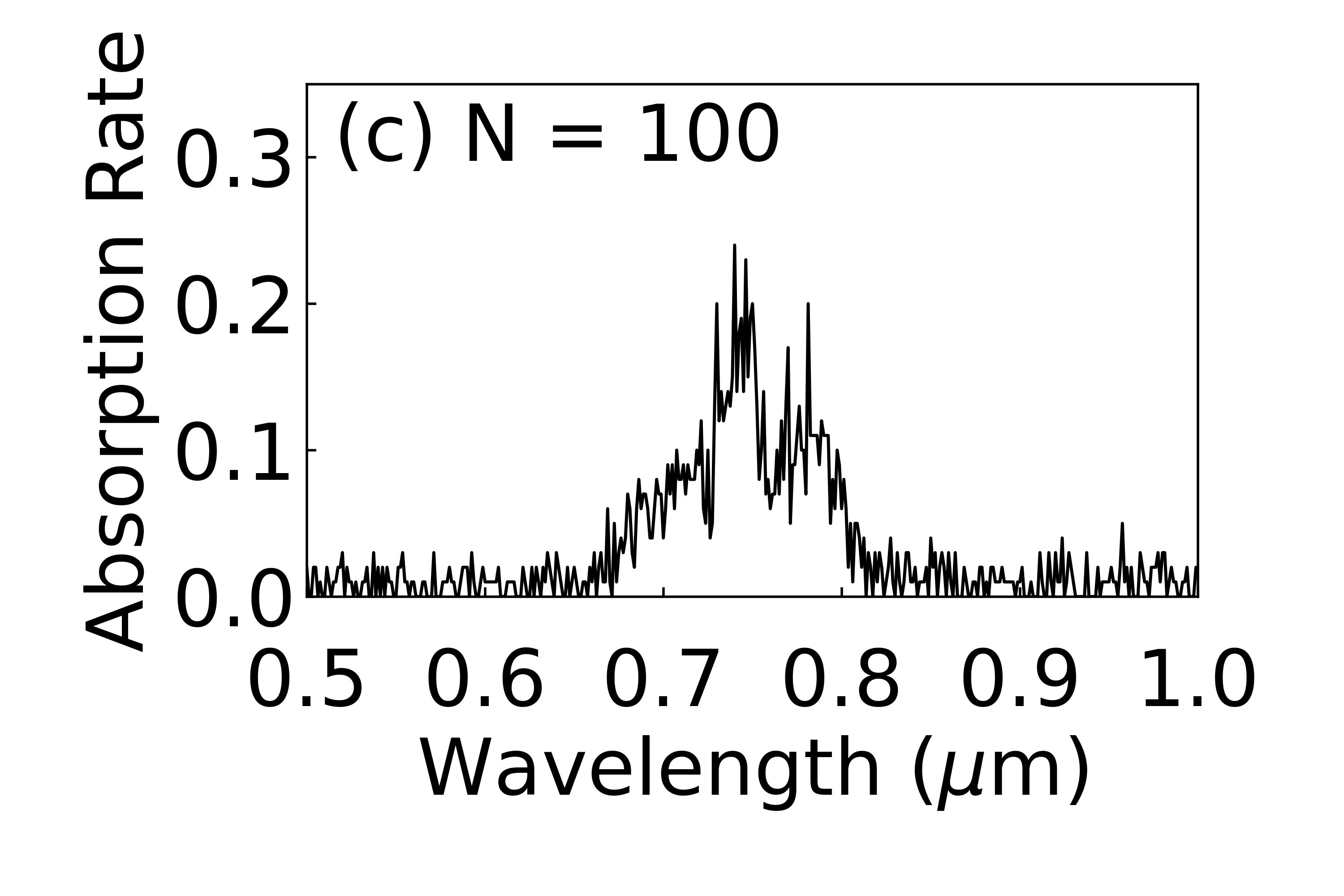"}
    \hspace{-0.7cm}
    \includegraphics*[width = .27\columnwidth]{"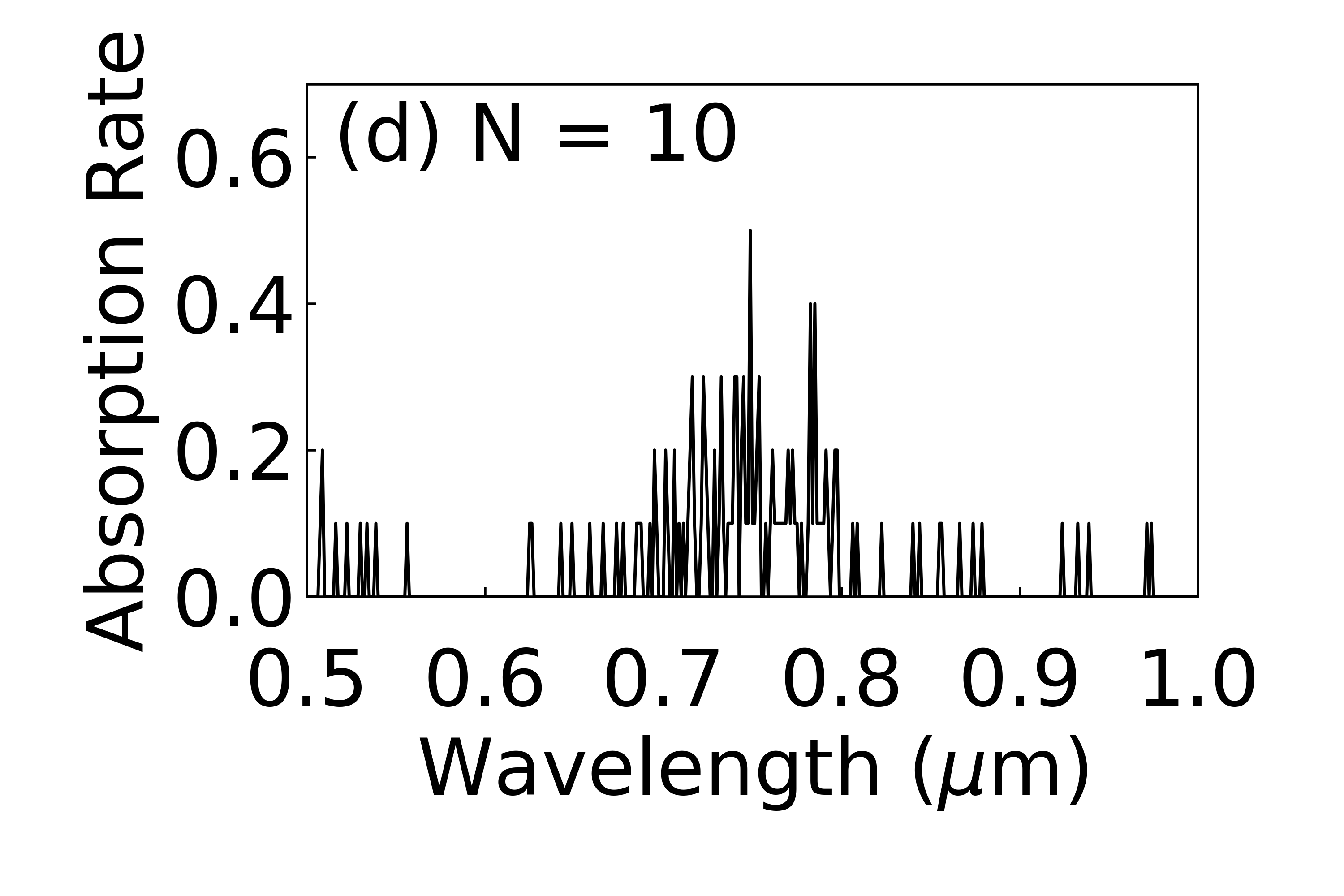"}\\
    \includegraphics*[width = .27\columnwidth]{"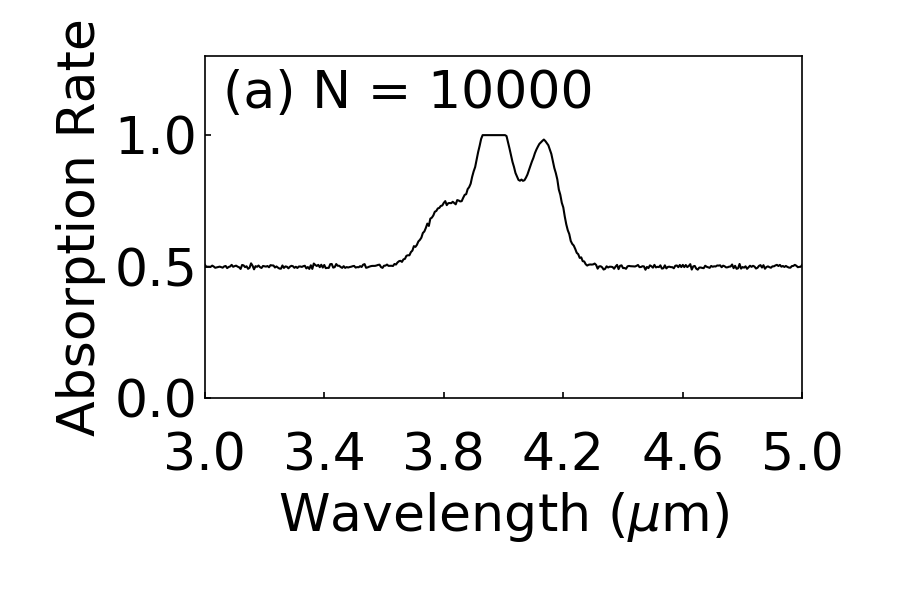"}
    \hspace{-0.7cm}
    \includegraphics*[width = .27\columnwidth]{"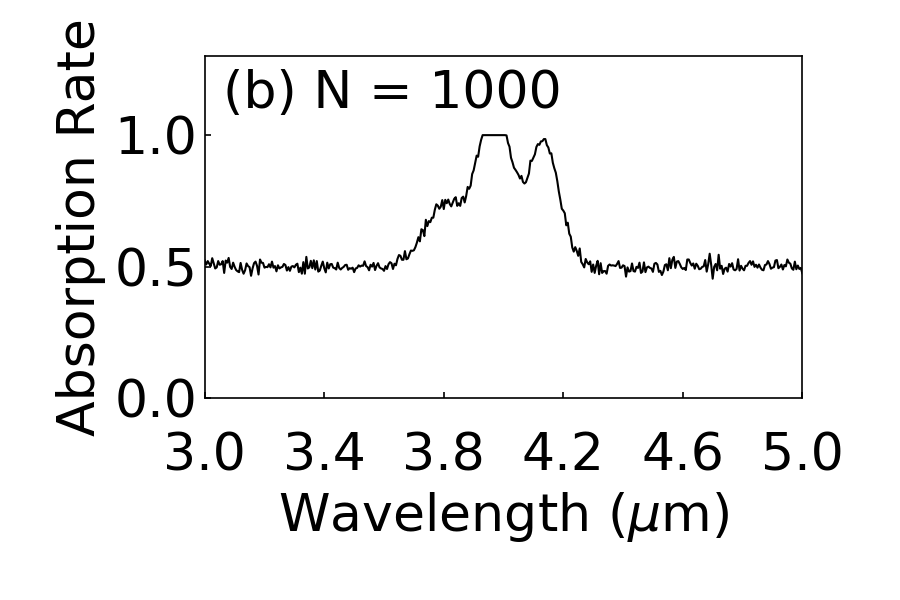"}
    \hspace{-0.7cm}
    \includegraphics*[width = .27\columnwidth]{"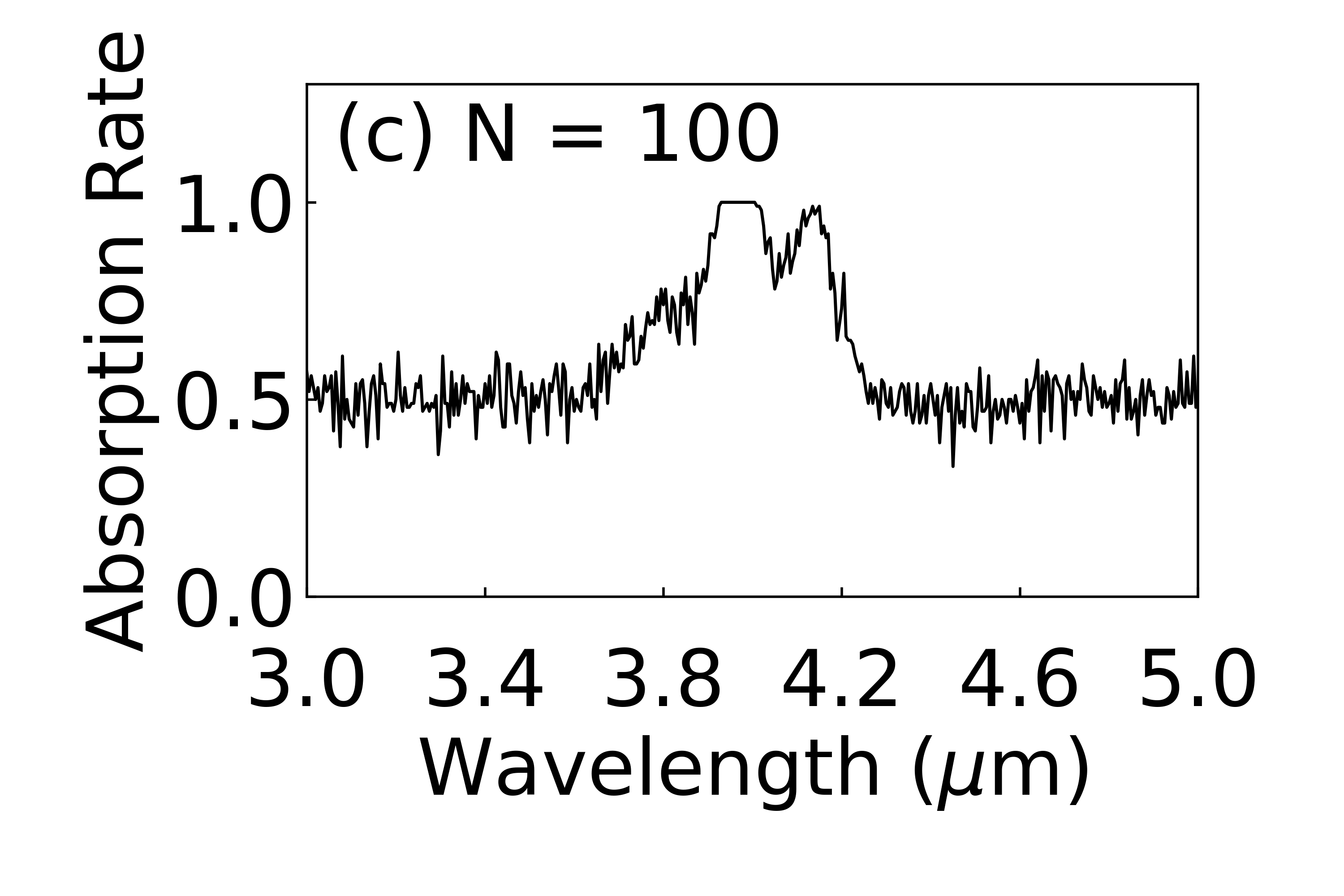"}
    \hspace{-0.7cm}
    \includegraphics*[width = .27\columnwidth]{"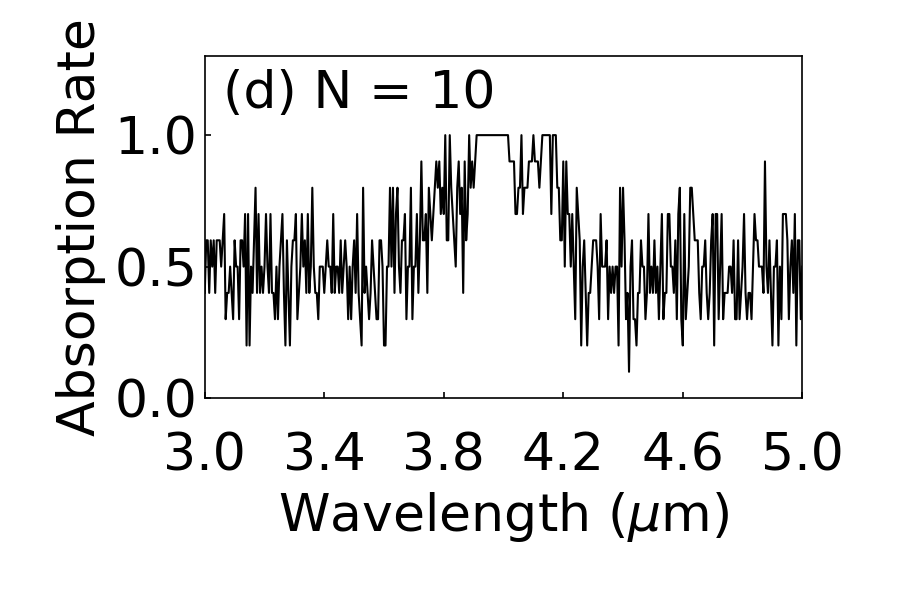"}
    \caption{Examples of artificial spectral data when peak intensities are observable and when peak intensities are not observable are shown on the upper and lower side respectively. Cases (a), (b), (c) and (d) are spectrum data when the number of incident photons per measurement point $N$ = 10000, 1000, 100 and 10, respectively.}
    \label{art_close_0}
\end{figure}

Here, for the number of incident photons per measurement point $N = 10000,1000,100$ and $10$, the data generated based on the basis of the binomial distribution are shown in the upper side of Fig. \ref{art_close_0}.
It can be seen that when the number of incident photons is small, the noise is large and it is difficult to determine the number of peaks.
Moreover, in this setting, the noise is large near the peaks. These properties are due to the fact that the variance of $n/N$ is $\alpha(1-\alpha)/N$ which is large when $\alpha$ is close to $0.5$ and when $N$ is small (where $N$ is the number of incident photons, $\alpha$ is the absorption rate, and $n$ is the number of absorbing photons.)\par
In this situation, let $\eta_a = 2.0, \lambda_a = 5.0,$ $\nu_0 = 0.5,\xi_0 = 1.0, \eta_\sigma = 10.0,\lambda_\sigma = 160.0,$ $\nu_B = 0.1,$ and $\xi_B = 0.01$, and the prior distributions of $\{a_k,\mu_k,\sigma_k\}_{k=1}^K,B$ be set as follows:
\begin{align}
    \label{a_prior}
    \varphi(a_k) &= \textup{Beta} \left(a_k;\eta_a,\lambda_a\right)\\ 
    &= \frac{a_k^{\eta_a -1}(1-a_k)^{\lambda_a-1}}{B(\eta_a,\lambda_a)}, \\
    \label{mu_prior} 
    \varphi(\mu_k) &= U(\nu_0,\nu_1), \\
    \label{sigma_prior}
    \varphi(\sigma_k) &= \textup{Gamma} \left(\frac{1}{\sigma_k^2};\eta_\sigma,\lambda_\sigma\right)\\
                      &= \frac{1}{\Gamma(\eta_b)}(\lambda_b)^{\eta_b}\left(\frac{1}{\sigma_k^2}\right)^{\eta_b-1}\exp\left(-\lambda_b \left(\frac{1}{\sigma_k^2}\right)\right), \\
    \label{B_prior}
    \varphi(B) &= N(B;\nu_B,{\xi_B}^2),
\end{align}
where $B(\eta_a,\lambda_a)$ is the beta function, $\Gamma(\eta_b)$ is the gamma function, $U(\nu_0,\nu_1)$ is the uniform distribution on $[\nu_0,\nu_1]$, and $N(B;\nu_B,{\xi_B}^2)$ is the Gaussian distribution of mean $\nu_B$ and variance ${\xi_B}^2$. \par
\subsubsection{Case where peak structures are not observable}
In the case where the peak structures are not observable, let the number of peaks $K=3$ and the true parameter $\theta^* = \{ \{ a_k^*, \mu_k^*, \sigma_k^* \}_{k=1}^3, B^* \}$ be as follows:
\begin{align}
    \begin{pmatrix}
        a_1^*\\
        a_2^*\\
        a_3^*\\
    \end{pmatrix} 
    = 
    \begin{pmatrix}
        0.2348\\
        0.6088\\
        0.4732\\
    \end{pmatrix}  ,\ 
    \begin{pmatrix}
        \mu_1^*\\
        \mu_2^*\\
        \mu_3^*\\
    \end{pmatrix}
    = 
    \begin{pmatrix}
        3.806\\
        3.970\\
        4.135\\
    \end{pmatrix} ,\ 
    \begin{pmatrix}
        \sigma_1^*\\
        \sigma_2^*\\
        \sigma_3^*\\
    \end{pmatrix}
    = 
    \begin{pmatrix}
        0.0682\\
        0.0550\\
        0.0520\\
    \end{pmatrix},\ 
    B = 0.5.
\end{align}
\color{black}
Assuming that the density of focused materials in the sample is small, we set a large background.
We quadrupled the peak intensities because when the target substance cannot be adjusted and the substance is measured over a wide range of wavelengths, the peak intensity can be large and all photons are absorbed near peaks.
These cases are possible in planetary science such as in the analysis of earth's atmosphere \cite{Goody1995}. 
\par
Here, for $N = 10000,1000,100$ and $10$, the data generated on the basis of the binomial distribution are shown in the lower side of Fig. \ref{art_close_0}.
These data have not been analyzed because the peak structures are not observable.
As for the noise, when the number of incident photons is small, the noise is large as in the small peak case. Moreover, in this cases, the variance of the data is small near peaks. These properties are due to the fact that the variance of absorption rate $\alpha$ is large when $\alpha$ is close to $0.5$ and small when $\alpha$ is closed to $0$ or $1$. 
In this situation, let $\eta_a = 5.0,\lambda_a = 0.1,$$\nu_0 = 3.0,\xi_0 = 5.0,\eta_\sigma = 10.0,\lambda_\sigma = 10.0,$$\nu_B = 0.5,$ and $\xi_B = 0.1$. The prior distributions of $\{a_k \}_{k=1}^K$ are set as follows:
\begin{align}
    \varphi(a_k) &= \textup{Gamma} \left(a_k;\eta_a,\lambda_a\right),
\end{align}
and the prior distributions of $\{\mu_k,\sigma_k\}_{k=1}^K, B$ are set as in Equation (\ref{mu_prior}),(\ref{sigma_prior}),(\ref{B_prior}).
\subsection{Bayesian Spectral Deconvolution Based on Binomial Distribution}
\begin{figure}[h]
    \centering
    \includegraphics*[width = .27\columnwidth]{"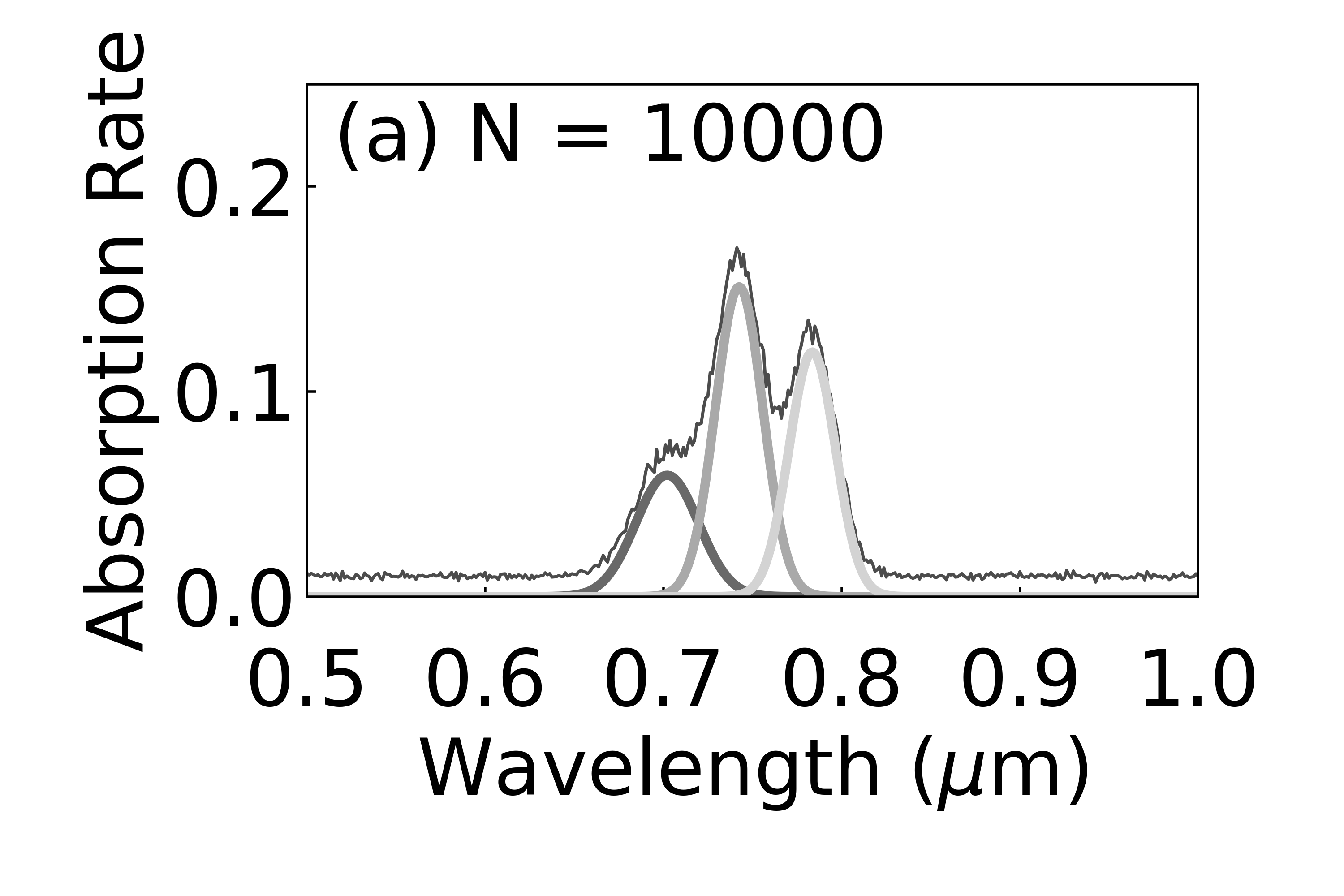"}
    \hspace{-0.7cm}
    \includegraphics*[width = .27\columnwidth]{"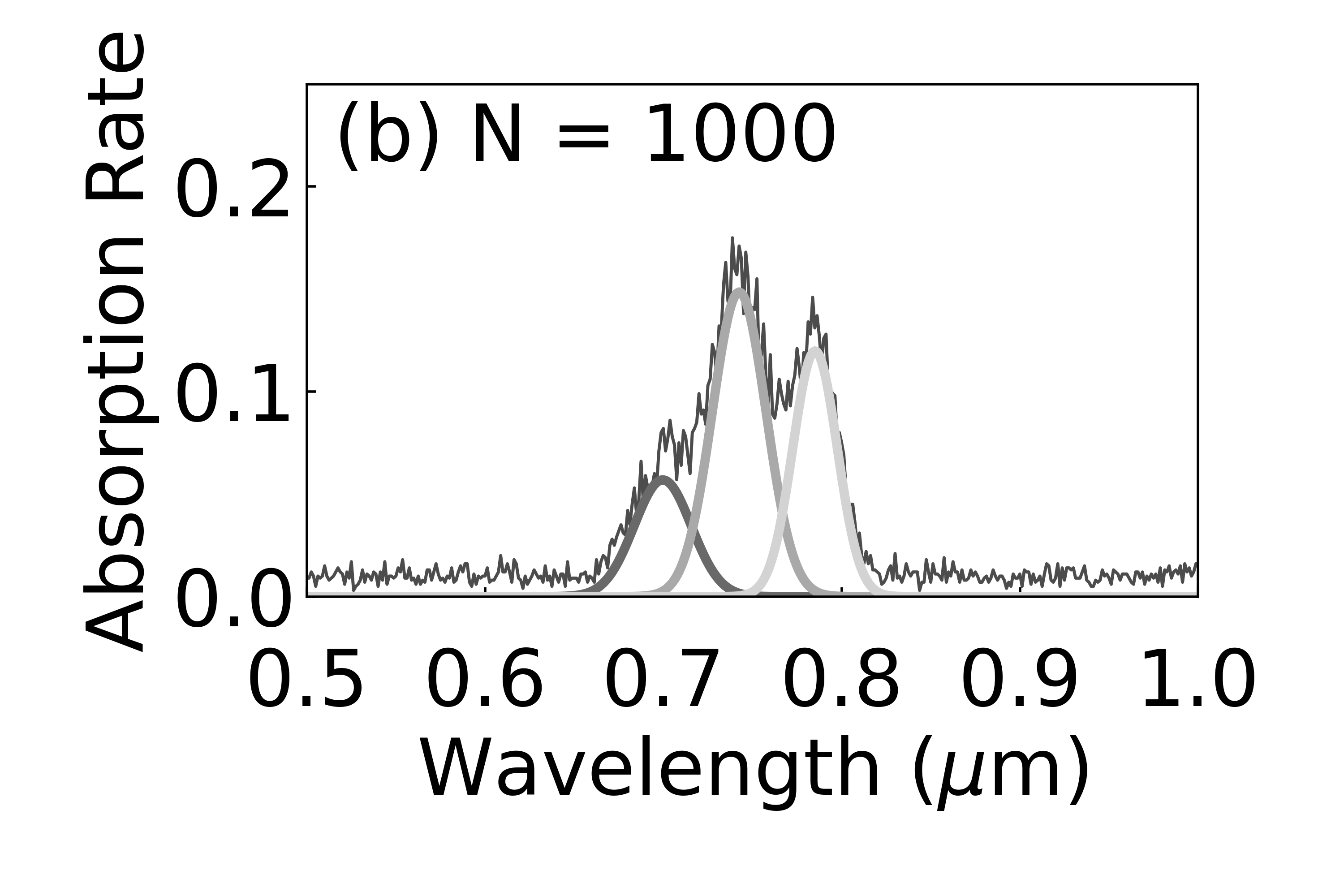"}
    \hspace{-0.7cm}
    \includegraphics*[width = .27\columnwidth]{"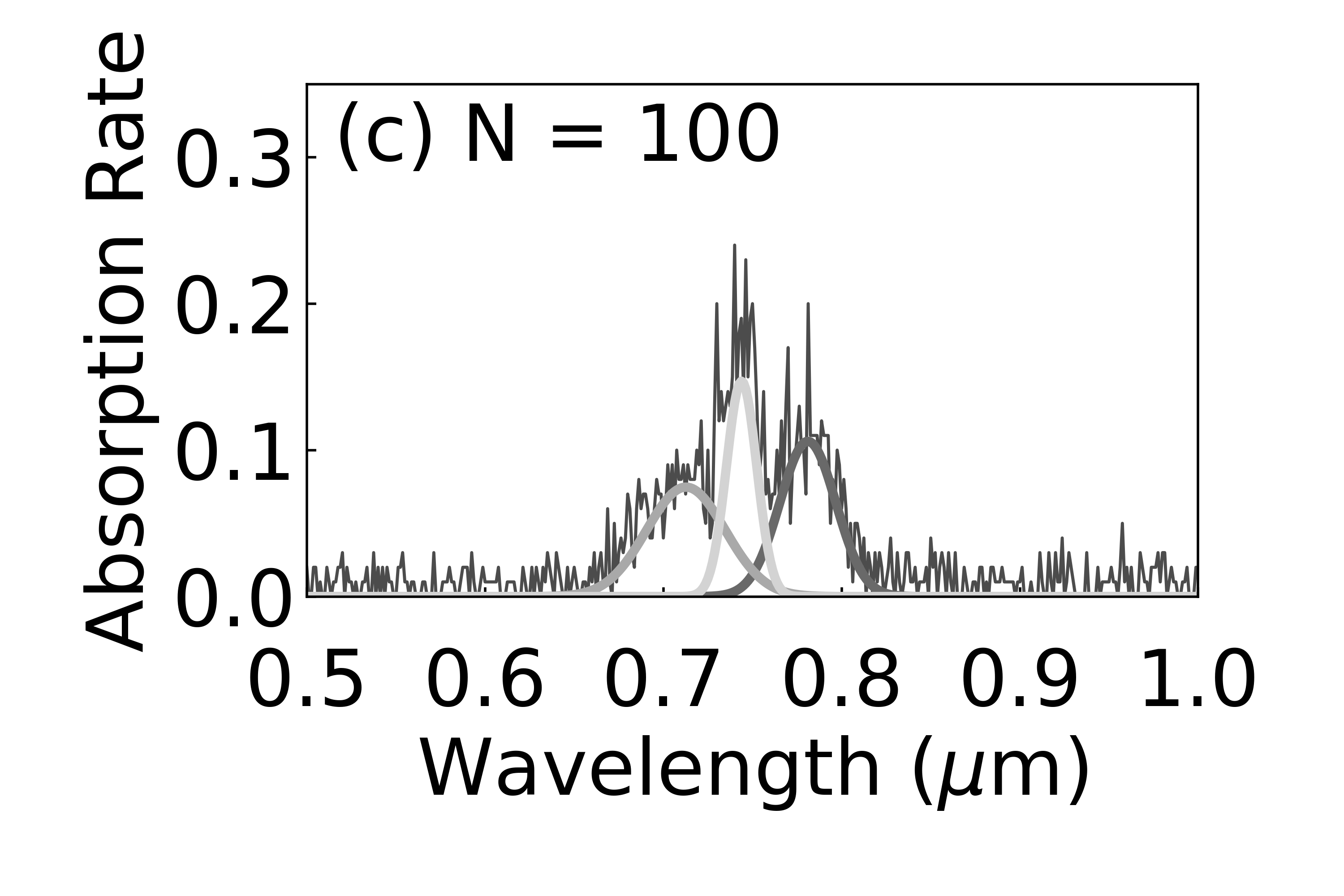"}
    \hspace{-0.7cm}
    \includegraphics*[width = .27\columnwidth]{"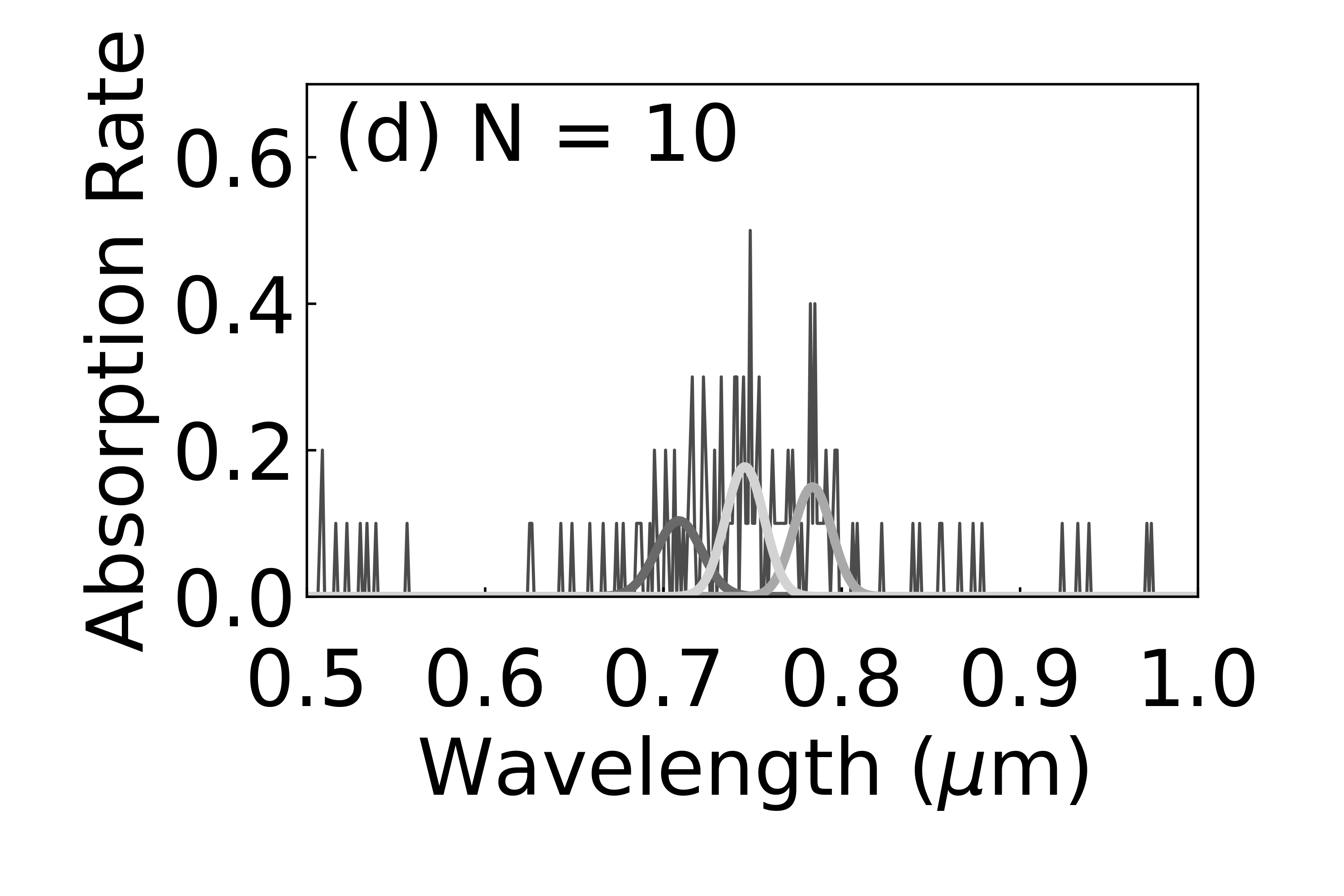"}
    \\
    \includegraphics*[width = .27\columnwidth]{"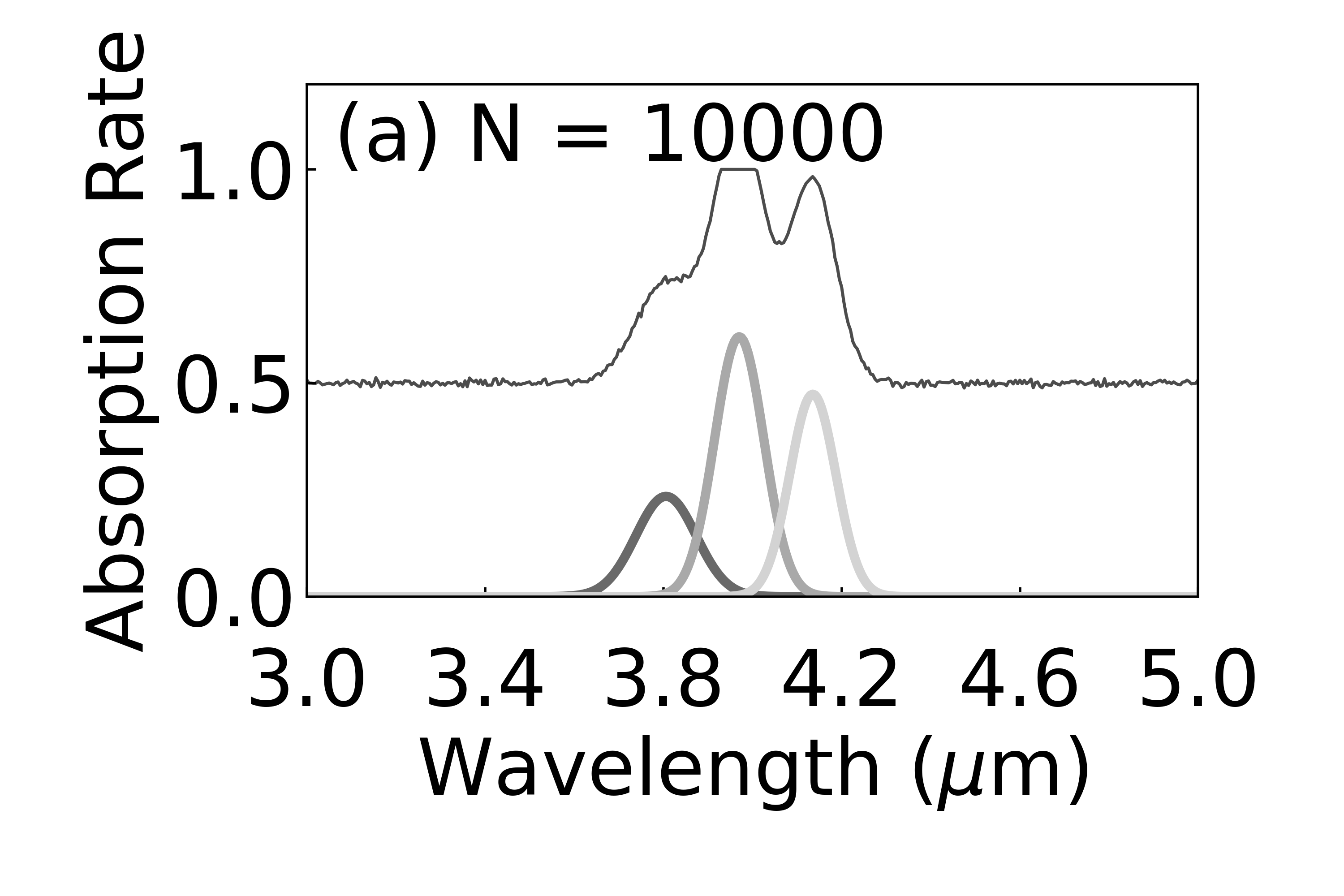"}
    \hspace{-0.7cm}
    \includegraphics*[width = .27\columnwidth]{"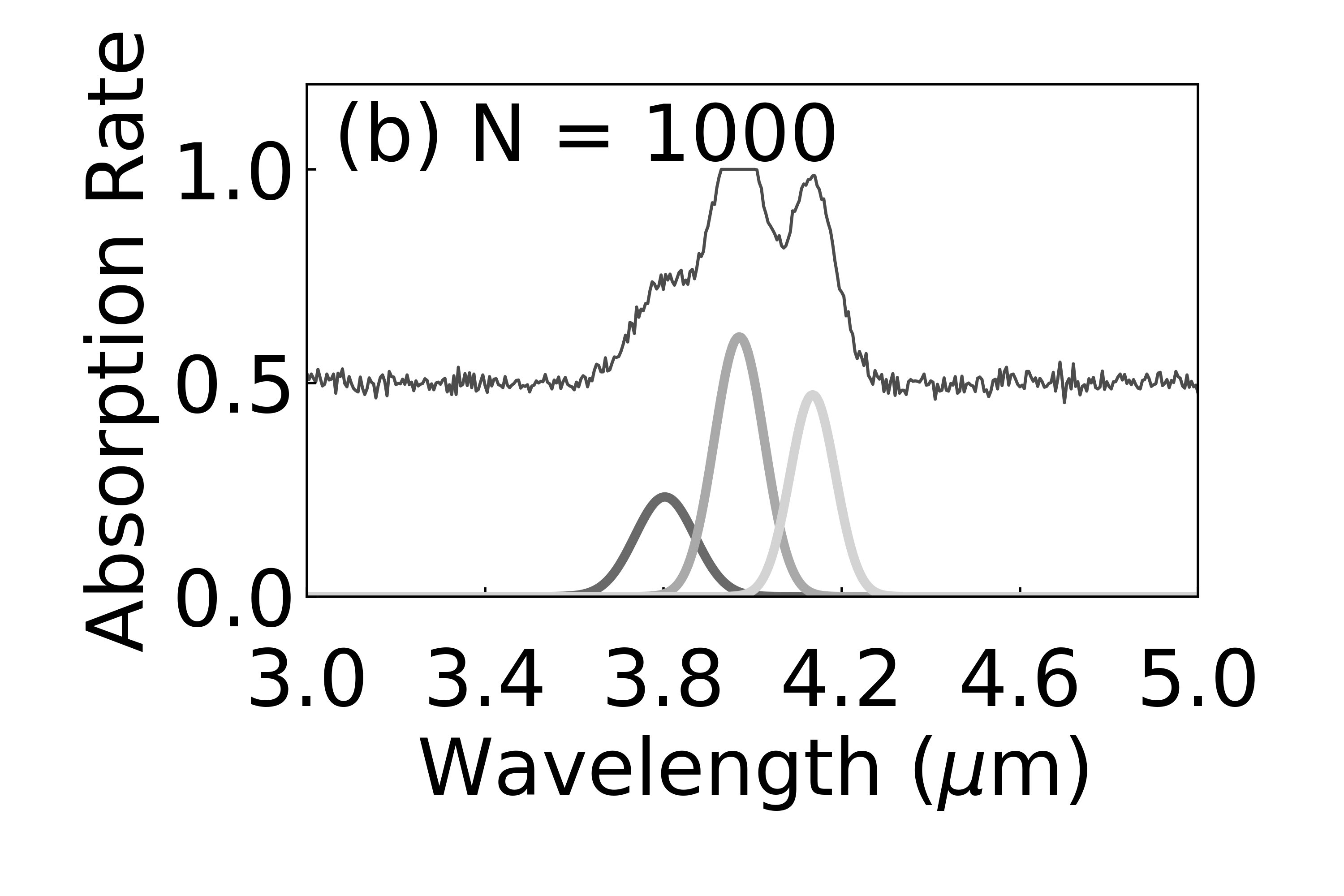"}
    \hspace{-0.7cm}
    \includegraphics*[width = .27\columnwidth]{"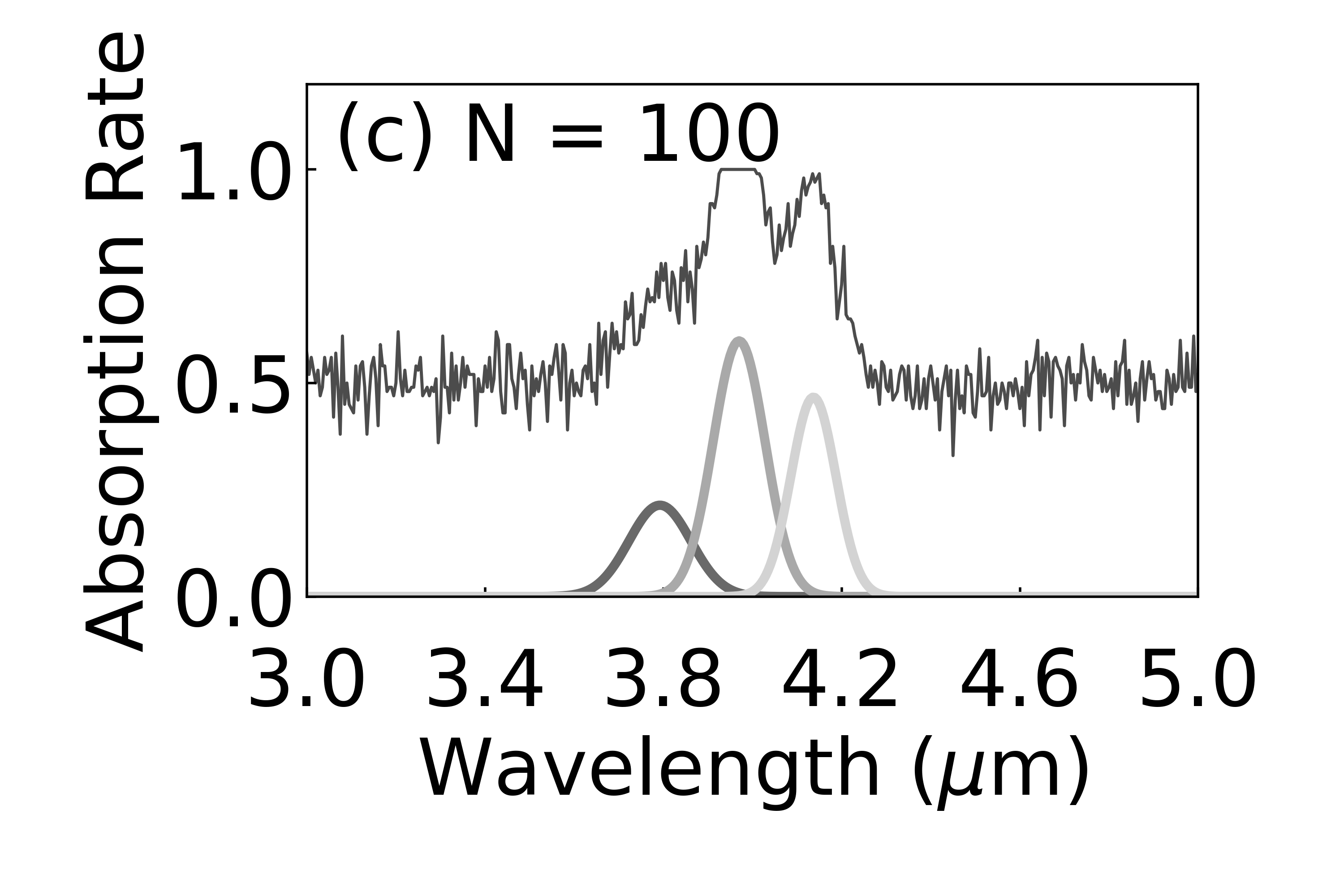"}
    \hspace{-0.7cm}
    \includegraphics*[width = .27\columnwidth]{"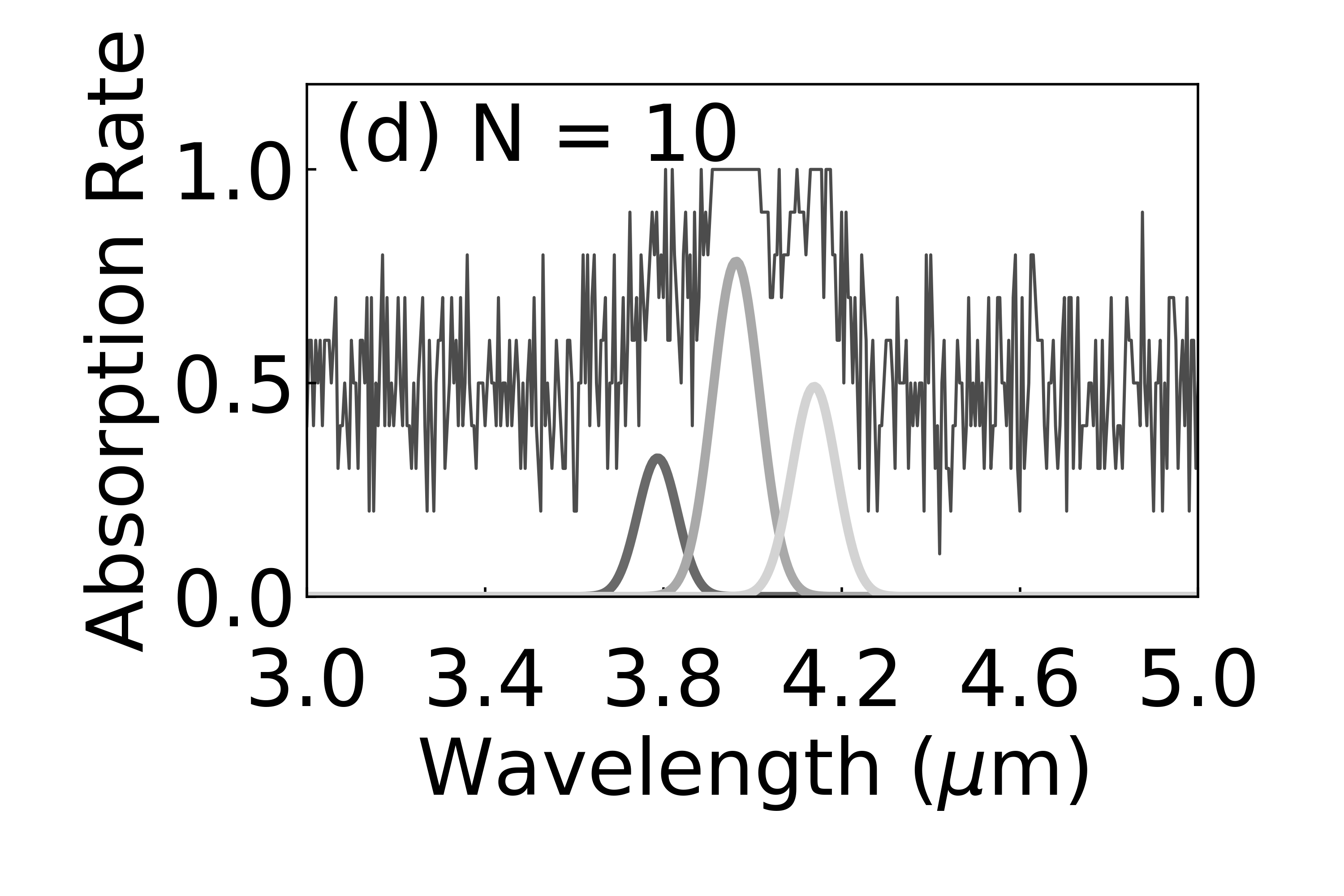"}
    \caption{Fittings for the estimated three peaks by Bayesian spectral deconvolution based on binomial distribution. The results when peak intensities are observable and when peak intensities are not observable are shown on the upper and lower side respectively. Cases (a), (b), (c) and (d) are spectrum data when the number of incident photons per measurement point $N$ = 10000, 1000, 100 and 10, respectively.}
    \label{fitting}
\end{figure}

\begin{figure}[h]
    \centering
    \includegraphics*[width = .27\columnwidth]{"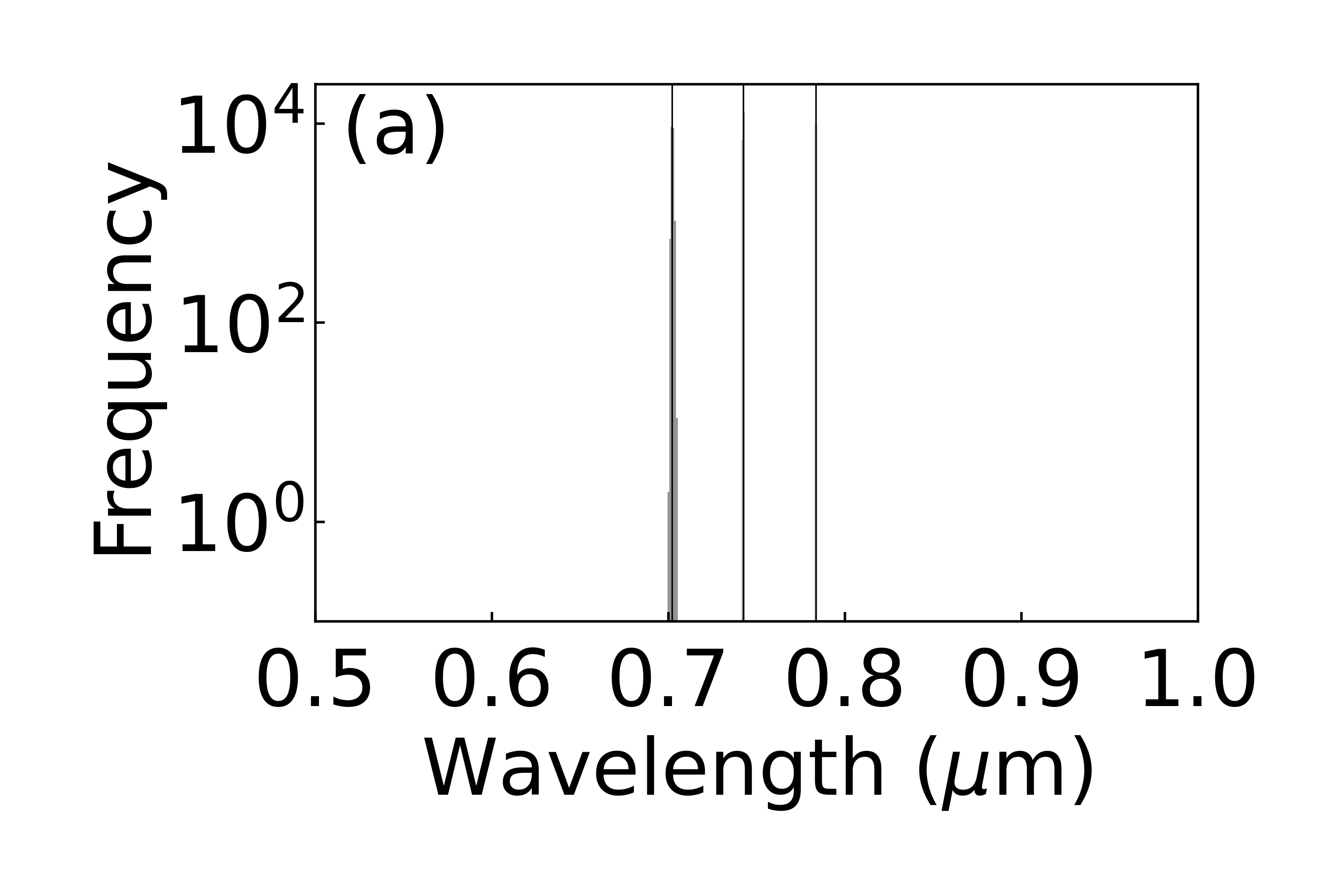"}
    \hspace{-0.7cm}
    \includegraphics*[width = .27\columnwidth]{"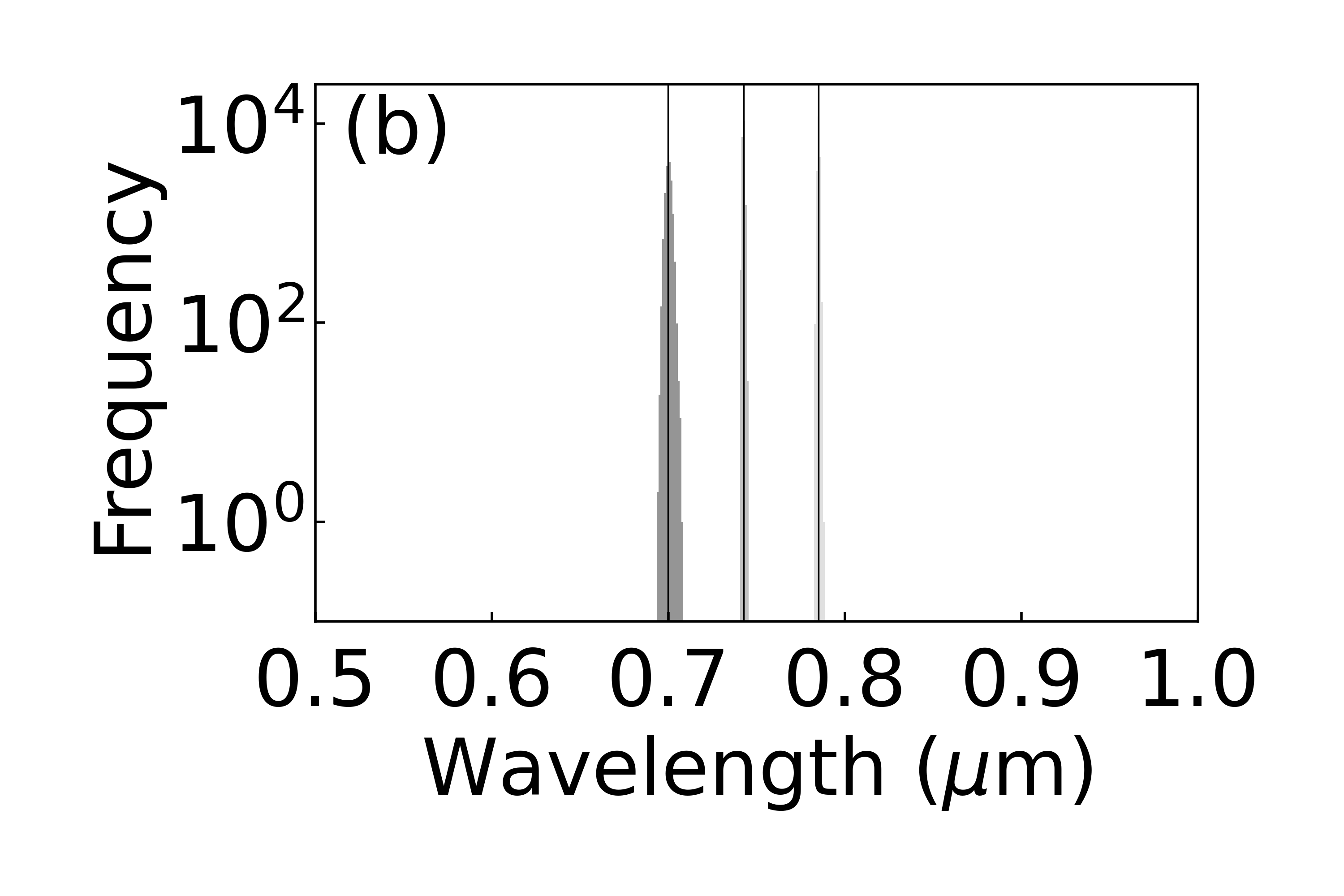"}
    \hspace{-0.7cm}
    \includegraphics*[width = .27\columnwidth]{"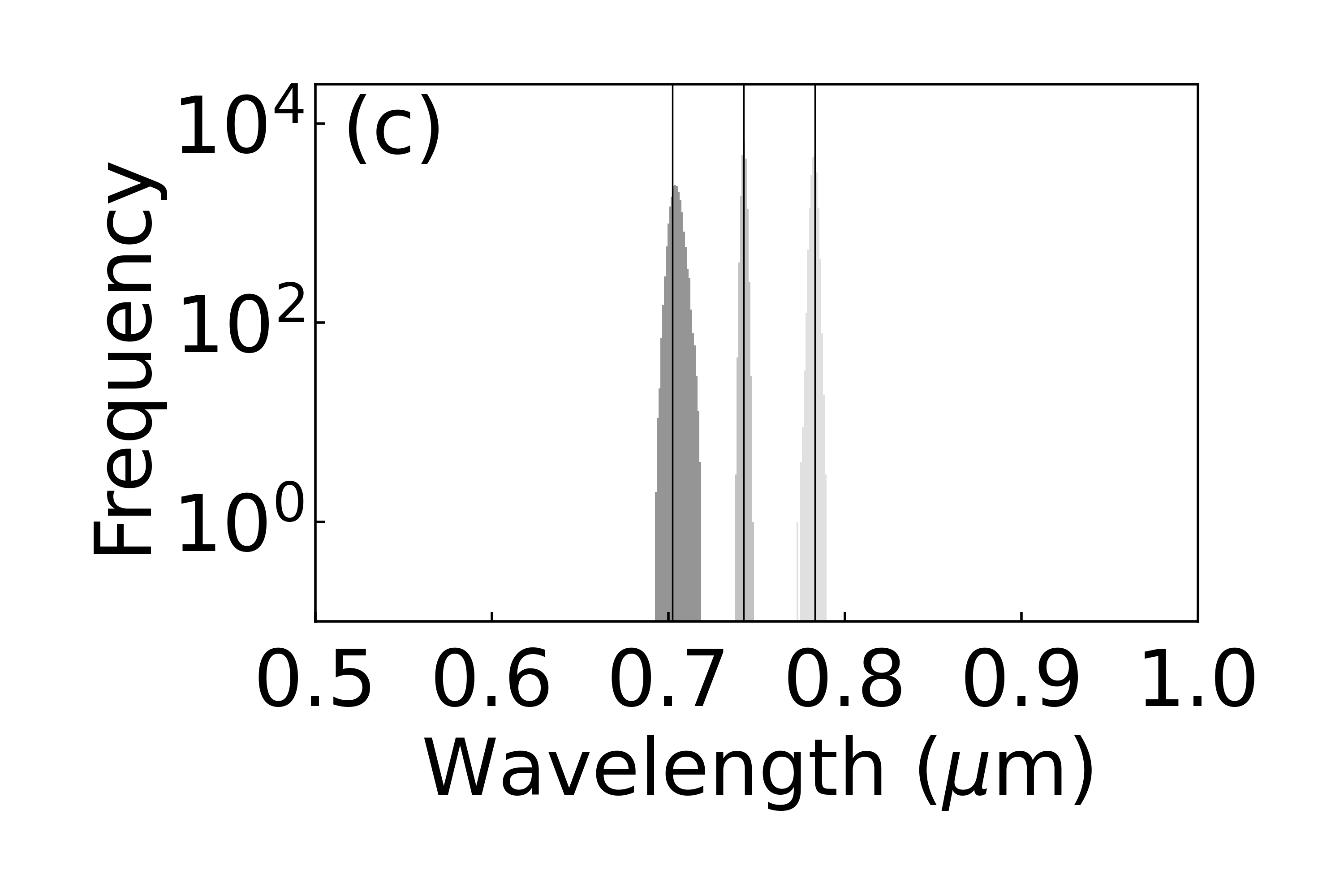"}
    \hspace{-0.7cm}
    \includegraphics*[width = .27\columnwidth]{"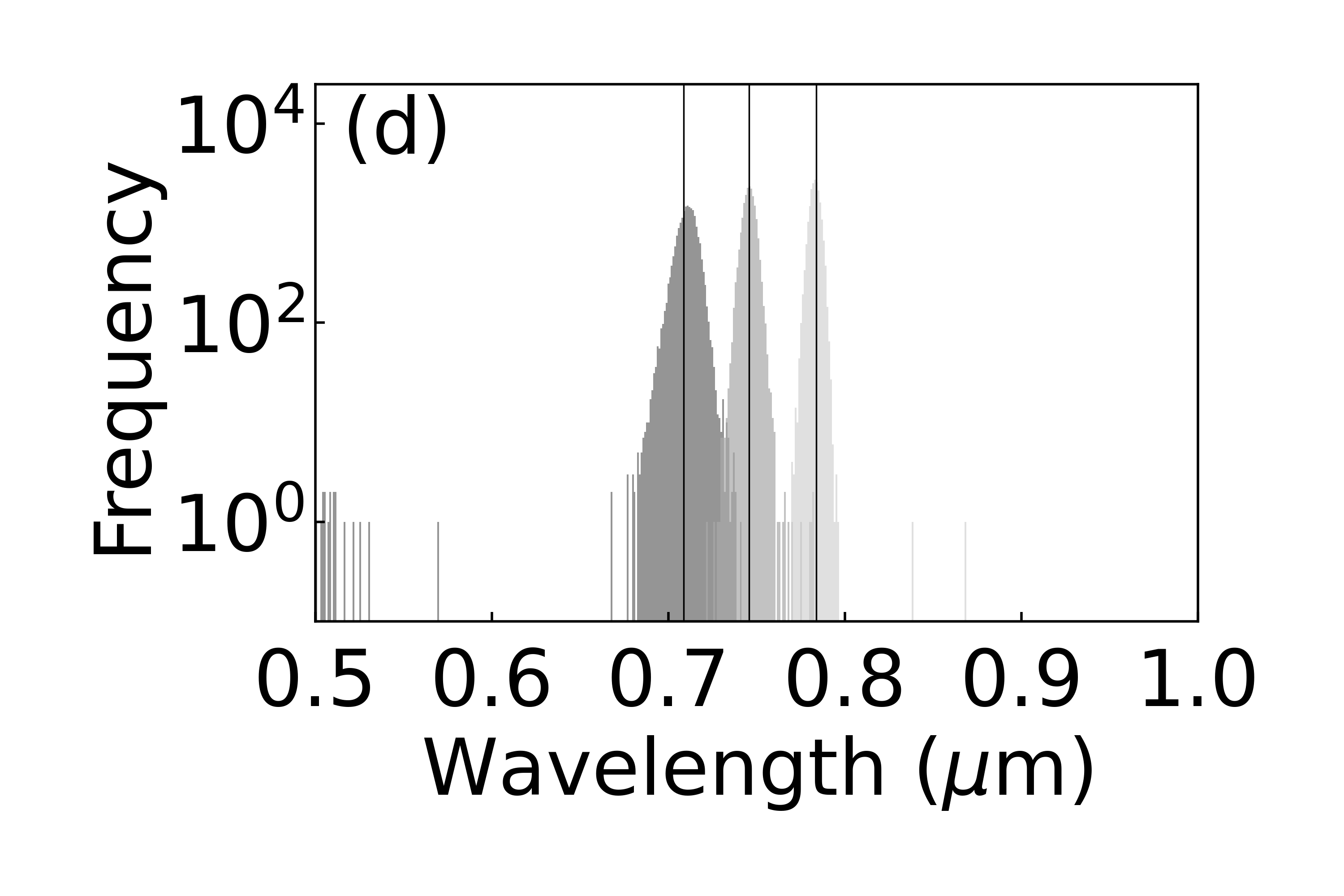"}\\
    \includegraphics*[width = .27\columnwidth]{"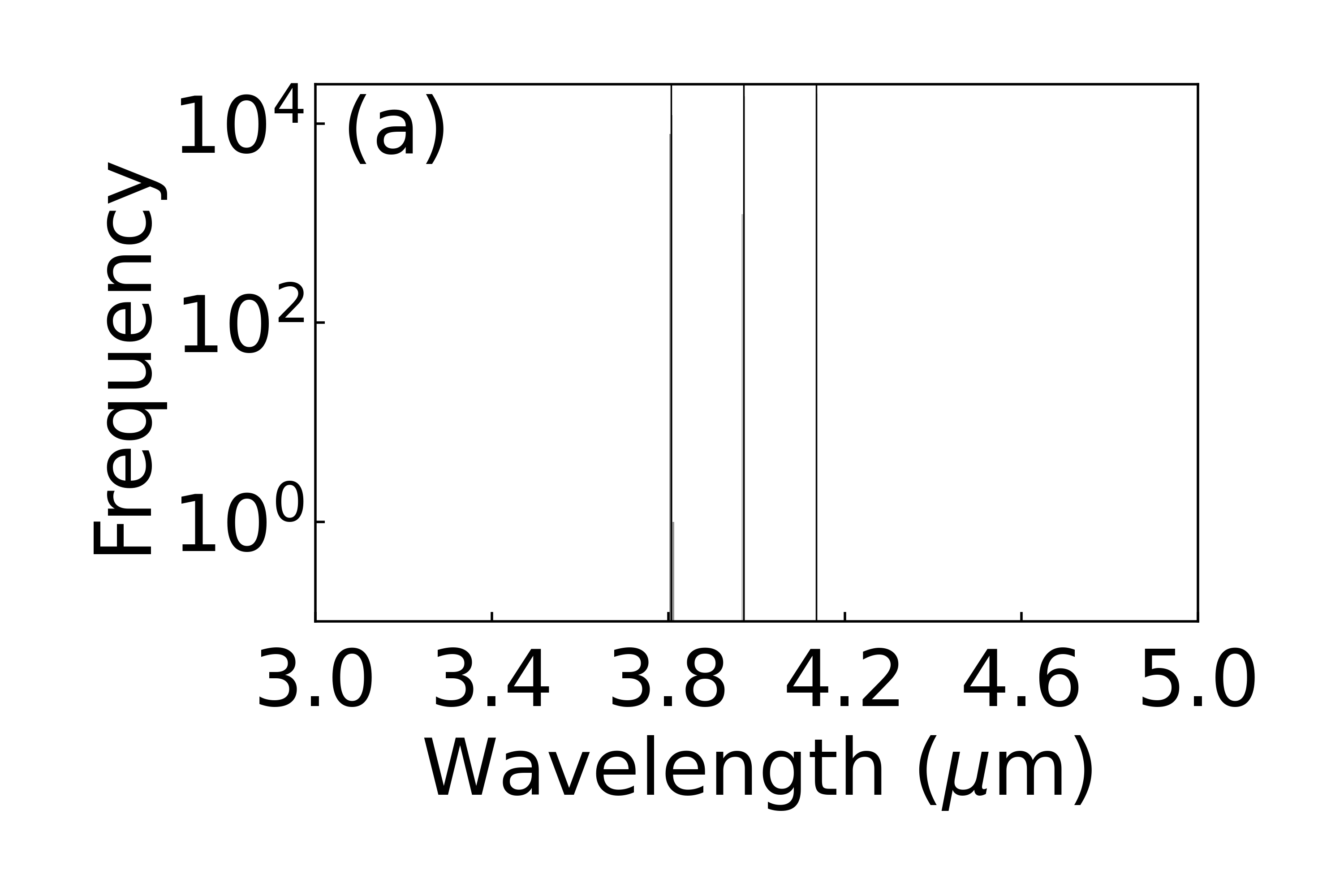"}
    \hspace{-0.7cm}
    \includegraphics*[width = .27\columnwidth]{"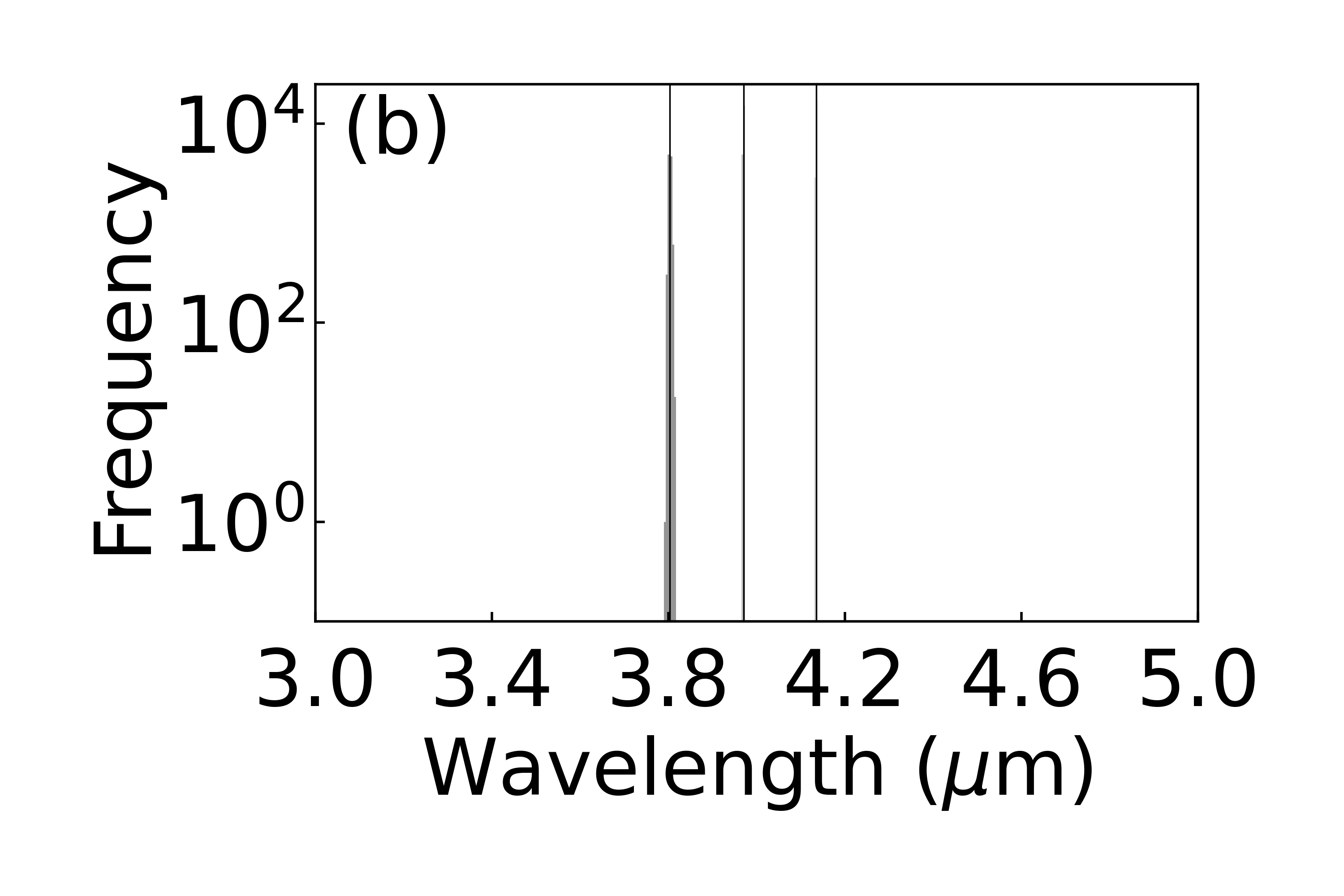"}
    \hspace{-0.7cm}
    \includegraphics*[width = .27\columnwidth]{"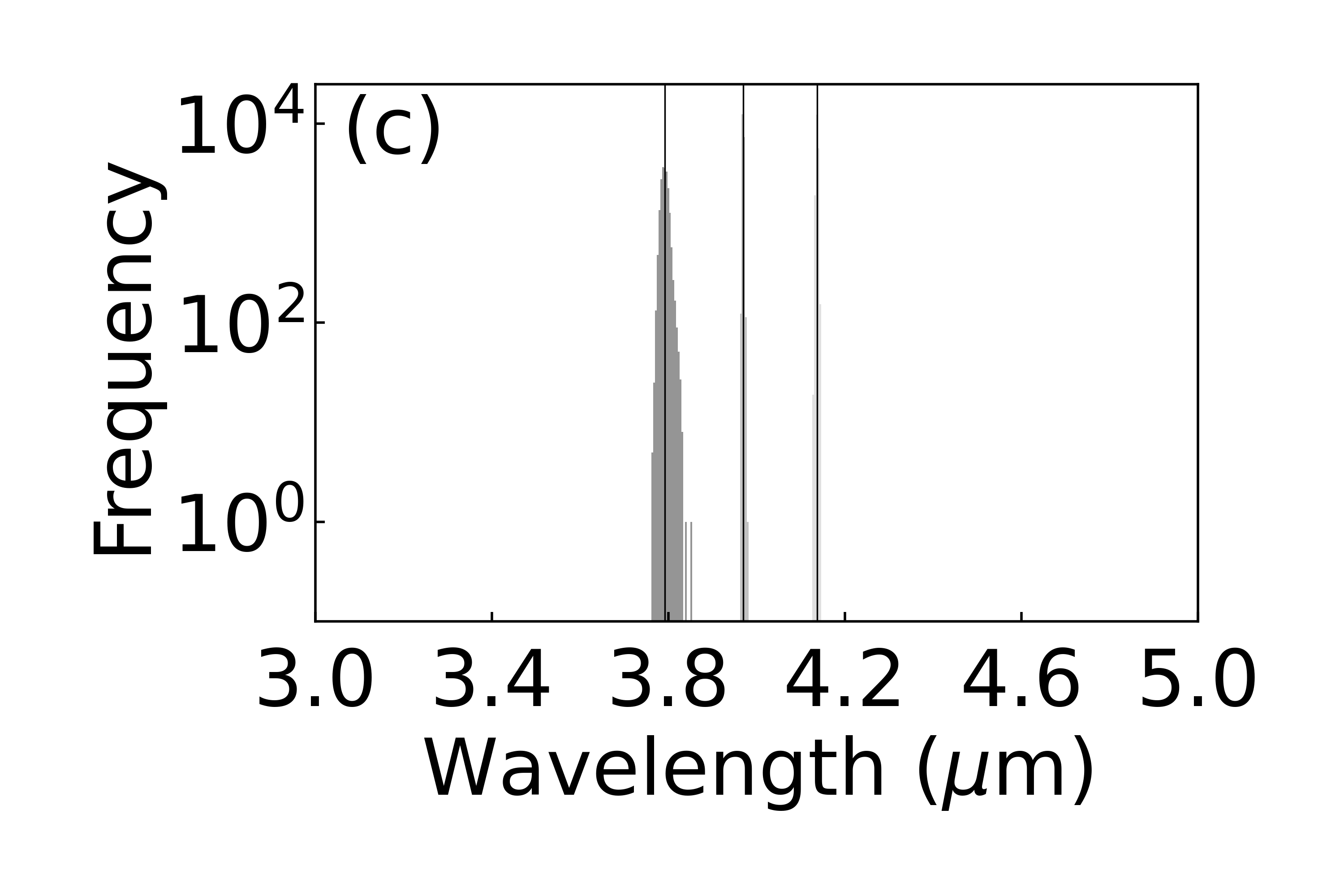"}
    \hspace{-0.7cm}
    \includegraphics*[width = .27\columnwidth]{"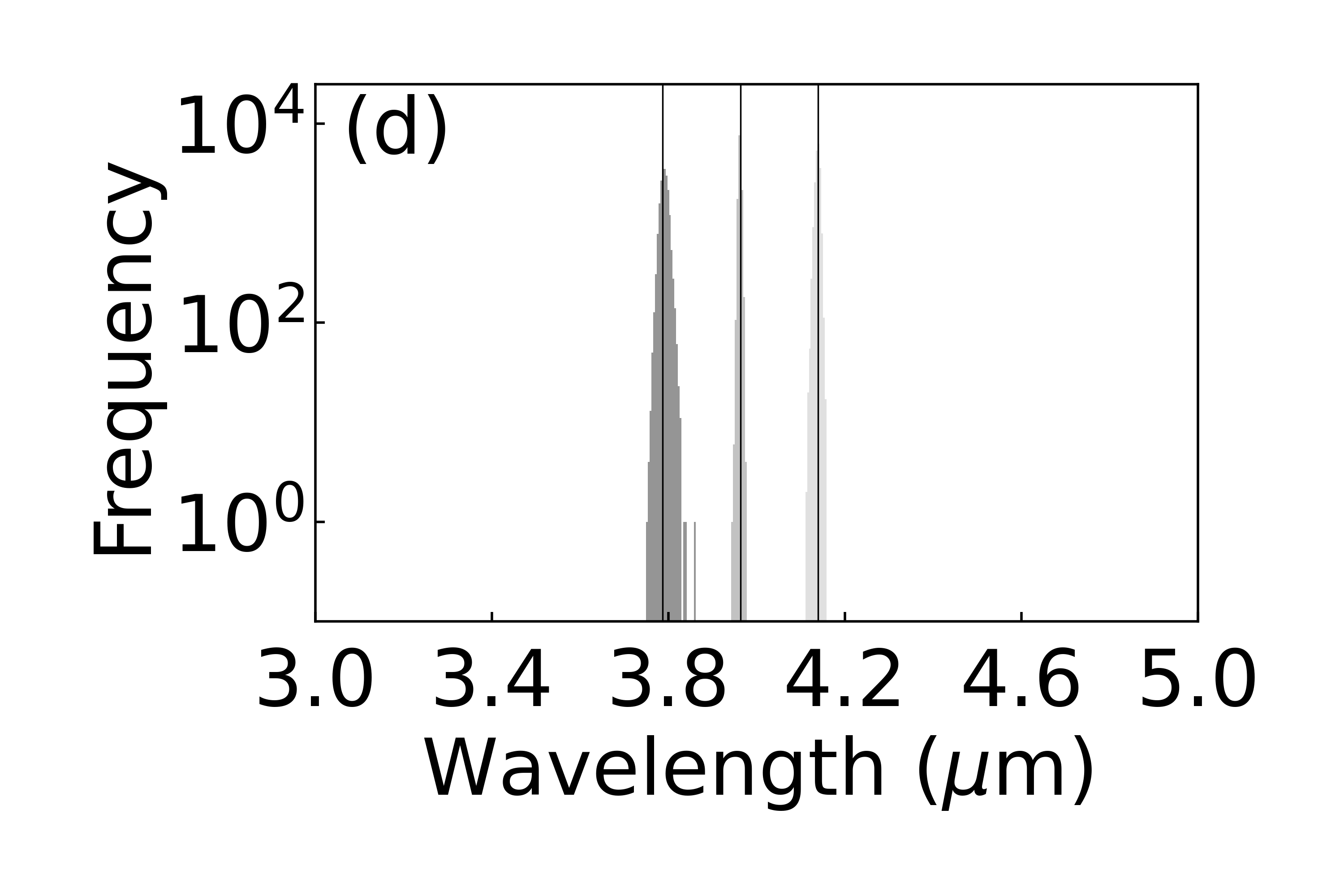"}

    \caption{The histograms of the posterior probability distribution $p(\mu_k|D,K)$ of the three peak positions $\mu_1,\mu_2,$ and $\mu_3$. The results when peak intensities are observable and when peak intensities are not observable are shown on the upper and lower side respectively. Cases (a), (b), (c) and (d) are spectrum data when the number of incident photons per measurement point $N$ = 10000, 1000, 100 and 10, respectively.}
    \label{parameter_estimation}
\end{figure}

Fittings of data by the Bayesian spectral deconvolution based on the binomial distribution are shown in Fig. \ref{fitting} and results of parameter estimations are shown in Fig. \ref{parameter_estimation}. 
It can be seen that the accuracy of parameters estimations improves as $N$ increases.
For the cases where peaks are observable and the case where peaks are not observable, parameters are separatedly distributed when $N = 100$ or more and when $N=10$ or more respectively, indicating that the parameter estimation is successfully performed.
In this study, we further generated 50 random spectral data for each cases, and we find the number of peaks $K$ that maximizes the posterior probability $p(K|D)$.
The results of model selection are shown in Table \ref{bin_cnt}.
\begin{table}[h]
    \caption{Frequency of model selection based on binomial distribution for each number of incident photons. The results when peak structures are observable and when peak structures are not observable are shown on the upper and the bottom side respectively.}
    \label{bin_cnt}
    \centering
     \begin{tabular}{lccccc}
      \hline
      Peak structures are observable & $K=1$ & $K=2$ & $K=3$ & $K=4$ & $K=5$ \\
      \hline 
      (a) $N = 10000$ &  0 & 0 & 50 & 0 & 0 \\
      (b) $N = 1000$ &  0 & 0 & 50 & 0 & 0 \\
      (c) $N = 100$ & 0 & 0&46 & 4 & 0  \\
      (d) $N = 10$ & 0& 2&42 & 5 & 1  \\
      \hline
     \end{tabular}\\
     \vspace{0.2cm}
     \begin{tabular}{lccccc}
        \hline
        Peak structures are not observable & $K=1$ & $K=2$ & $K=3$ & $K=4$ & $K=5$ \\
        \hline 
        (a) $N = 10000$ &  0 & 0 & 50 & 0 & 0 \\
        (b) $N = 1000$ &  0 & 0 & 49 & 1 & 0 \\
        (c) $N = 100$ & 0 & 0& 34 & 15 & 1  \\
        (d) $N = 10$ & 0 & 0 & 5 & 38 & 7  \\
        \hline
       \end{tabular}
   \end{table}
For the cases where peak structures are observable and where peak structures are not observable, the correct model $K = 3$ is selected in all cases when $N = 1000$ or more and when $N=10000$ respectively, indicating that the model selection is successful.
\section{Conclusion and Future Work}
In this study, we have developed a Bayesian spectral deconvolution method for absorption spectral data, the noise of which can be assumed to follow a binomial distribution.
Using artificial data, we perform Bayesian spectral deconvolution in the cases where absorption rate is small enough to observe the peak structures and too large to observe the peak structures.
We showed that we can estimate parameters and the number of peaks with high accuracy on the data where the incident photon is small and where spectral structure is flattened because of large peaks.\par 
Moreover, our method is also expected to be applicable to cases where the number of incident photons varies.
In actual experiments, the number of incident photons fluctuates in time and also changes with the energy of the incident light.
In the analysis of absorption spectra, the change in the number of incident photons has not been considered. For example, in the NASA/Keck
RELAB database located at Brown University \cite{RELAB}, the number of incident photons has not been stored.
To apply our method, not only the absorption rate but also the number of incident photons have to be stored in database.
\par
In our future work, we will apply our method to real data such as XAS and IR.
However, in general, there are no accurate forward models for XAS data, which makes model selection and parameter estimation difficult.
Thus, it is important to know what to do when phenomena cannot be represented by a forward model.

\end{document}